%% file: main.tex
\documentclass[manuscript, screen]{acmart}
\AtBeginDocument{%
  \providecommand\BibTeX{{%
    \normalfont B\kern-0.5em{\scshape i\kern-0.25em b}\kern-0.8em\TeX}}}

\usepackage{booktabs} 

\usepackage[english]{babel}
\usepackage{moresize}
\usepackage{amsmath}
\usepackage{algorithmic}
\usepackage{balance}
\usepackage{comment}
\usepackage{paralist}
\usepackage{bm}
\usepackage{pgfplots}
\usetikzlibrary{pgfplots.dateplot}

\usepackage{fancyhdr}
\usepackage{flushend}
\usepackage[english]{babel}
\usepackage[latin1]{inputenc}
\usepackage{mathrsfs}
\usepackage{graphicx}

\usepackage{amssymb}
\usepackage{amsfonts}
\usepackage{url}
\usepackage{longtable}
\usepackage{rotating}
\usepackage{multirow}
\usepackage{mathrsfs}
\usepackage{subfigure}
\usepackage{enumitem}
\usepackage[linesnumbered,algoruled,boxed,lined]{algorithm2e}
\usepackage{adjustbox}
\usepackage{hyperref}
\usepackage{pgfplots}
\usetikzlibrary{pgfplots.dateplot}
\usepackage{filecontents}
\definecolor{tblue}{RGB}{31,119,180}
\definecolor{torange}{RGB}{255,127,14}
\definecolor{tgreen}{RGB}{44,160,44}
\definecolor{tred}{RGB}{214,39,40}
\definecolor{tpurple}{RGB}{148,103,189}

\newcommand{\hide}[1]{} 

\newcommand{\ie}{\textit{i}.\textit{e}.}
\newcommand{\eg}{\textit{e}.\textit{g}.}

\setcopyright{acmlicensed}
\acmJournal{CSUR}
\acmYear{2025} \acmVolume{1} \acmNumber{1} \acmArticle{1} \acmMonth{1}\acmDOI{10.1145/3746280}

\begin{document}

\title{A Comprehensive Survey on Self-Supervised Learning for Recommendation}

\author{Xubin Ren}
\affiliation{%
  \institution{The University of Hong Kong}
  \city{Hong Kong}
  \country{China}}
\email{xubinrencs@gmail.com}
  
\author{Wei Wei}
\affiliation{%
  \institution{The University of Hong Kong}
  \city{Hong Kong}
  \country{China}}
\email{weiweics@connect.hku.hk}

\author{Lianghao Xia}
\affiliation{%
  \institution{The University of Hong Kong}
  \city{Hong Kong}
  \country{China}}
\email{aka_xia@foxmail.com}

\author{Chao Huang}
\authornote{Corresponding author}
\affiliation{%
  \institution{The University of Hong Kong}
  \city{Hong Kong}
  \country{China}}
\email{chaohuang75@gmail.com}


\begin{abstract}
Recommender systems play a crucial role in tackling the challenge of information overload by delivering personalized recommendations based on individual user preferences. Deep learning techniques, such as RNNs, GNNs, and Transformer architectures, have significantly propelled the advancement of recommender systems by enhancing their comprehension of user behaviors and preferences. However, supervised learning methods encounter challenges in real-life scenarios due to data sparsity, resulting in limitations in their ability to learn representations effectively. To address this, self-supervised learning (SSL) techniques have emerged as a solution, leveraging inherent data structures to generate supervision signals without relying solely on labeled data. By leveraging unlabeled data and extracting meaningful representations, recommender systems utilizing SSL can make accurate predictions and recommendations even when confronted with data sparsity. In this paper, we provide a comprehensive review of self-supervised learning frameworks designed for recommender systems, encompassing a thorough analysis of over 170 papers. We conduct an exploration of nine distinct scenarios, enabling a comprehensive understanding of SSL-enhanced recommenders in different contexts. For each domain, we elaborate on different self-supervised learning paradigms, namely contrastive learning, generative learning, and adversarial learning, so as to present technical details of how SSL enhances recommender systems in various contexts. We consistently maintain the related open-source materials at \color{blue}{\url{https://github.com/HKUDS/Awesome-SSLRec-Papers}}.
\end{abstract}

\ccsdesc[500]{General and reference~Surveys and overviews}
\ccsdesc[500]{Information systems~Recommender systems}

\keywords{Recommendation, Self-Supervised Learning}


\maketitle

\input{intro}
\input{preli}
\input{taxon}
\input{scena}
\input{future}
\input{conclusion}

\bibliographystyle{ACM-Reference-Format}
\balance
\bibliography{sample-base}
\clearpage

\end{document}

%% file: intro.tex
\section{Introduction}
\label{sec:intro}

Recommender systems play a vital role in addressing the challenge of information overload by providing personalized recommendations to individual users based on their unique preferences~\cite{zhang2019deep}. These systems are designed to enhance the overall user experience by presenting users with recommendations that are not only relevant but also closely aligned with their interests. This tailored approach makes the user experience more engaging, efficient, and ultimately more satisfying. At the core of recommender systems lies the fundamental principle of understanding users' preferences for a diverse range of items~\cite{SimpleX,DENS}. This understanding is achieved through a meticulous analysis of users' past interactions, which encompass activities such as clicks and purchases. By examining these interactions, recommender systems gain valuable insights into users' behavior, enabling them to identify patterns and uncover individual preferences.

Recommender systems have undergone revolutionary transformation through deep learning techniques. Neural architectures including RNNs \cite{jannach2017recurrent,guo2020attentional}, GNNs \cite{ying2018graph,wang2019neural}, and Transformers \cite{BERT4Rec,wu2020sse} have enabled unprecedented understanding of user preferences, facilitating precise personalized recommendations. However, existing supervised learning methods require abundant labeled data, while practical recommender systems often face data sparsity~\cite{HCCF}, limiting effective generalization and accurate preference learning. Inspired by self-supervised learning (SSL) successes~\cite{liu2021self}, SSL techniques have proven beneficial for addressing data sparsity in recommender systems~\cite{DCCF, HCCF}. SSL leverages inherent data structures to create supervision signals without relying on external labels, enabling systems to utilize unlabeled data and extract meaningful representations for accurate predictions despite data sparsity.

Our paper provides a comprehensive review of the latest advancements in self-supervised learning frameworks tailored specifically for recommendation systems. It aims to serve as a valuable resource for researchers from diverse disciplines, extending beyond the realms of computer science and machine learning, who wish to explore this rapidly evolving field. In this context, our paper presents several significant contributions, which are summarized as follows: \\\vspace{-0.12in}

\noindent \textbf{Comprehensive Collection of Papers.} We have conducted a thorough review of over 170 papers that investigate the application of self-supervised learning in the field of recommendation. Our search was carried out using reputable academic databases such as Google Scholar and DBLP, utilizing specific keywords including ``self-supervised'', ``contrastive'', ``generative'', ``adversarial'', ``variational'', ``diffusion'' and ``masked autoencoder'' in conjunction with ``recommendation'' and ``recommender systems''. The surveyed papers were meticulously sourced from esteemed conferences and journals such as KDD, SIGIR, WWW, ICLR, WSDM, CIKM, ICDE, AAAI, IJCAI, RecSys, TOIS, TKDE. To ensure the inclusion of state-of-the-art research, we also explored citation networks and incorporated relevant preprints from arXiv. \\\vspace{-0.12in}

\noindent \textbf{Supported with Open-source Library.} Our team has developed SSLRec~\cite{SSLRec}, a robust framework for self-supervised learning. This user-friendly framework encompasses popular datasets, standardized code scripts for data processing, training, testing, evaluation, and cutting-edge recommender models for self-supervised learning. \\\vspace{-0.12in}

\noindent \textbf{Relationship with Previous Surveys.} i) \textbf{Extensive Collection of SSL Research}. Our survey provides a significantly broader collection of self-supervised learning (SSL) works in the context of recommendation, encompassing over 170 papers, surpassing the previous surveys which covered approximately 80 papers~\cite{yu2023self, jing2023contrastive}. In addition, we incorporate reconstructive methods (e.g., mask autoencoding, denoised diffusion) and adversarial learning, which were not previously covered but have gained prominence according to~\cite{liu2021self}. ii) \textbf{Comprehensive Taxonomy Design}. Our taxonomy refines previous classifications of contrastive learning~\cite{yu2023self, jing2023contrastive} and proposes a view-centric taxonomy approach. We also survey reconstructive and adversarial learning methods based on their learning paradigms and targets. iii) \textbf{Exploration of Distinct Scenarios}. Distinguishing ourselves from earlier works~\cite{yu2023self, jing2023contrastive}, which did not differentiate SSL-enhanced recommenders across different scenarios, we conducted an extensive exploration of nine scenarios individually. By surveying research works within diverse scenarios, we provide researchers with a more comprehensive understanding of the context-specific to each scenario and the corresponding challenges they present. \\\vspace{-0.12in}

\noindent \textbf{Organization of the Survey:} The structure of the survey is as follows: Section~\ref{sec:preli} provides a comprehensive overview of recommendation systems and self-supervised learning, establishing the necessary background knowledge. In Section~\ref{sec:taxonomy}, we present our proposed taxonomy for understanding self-supervised learning within the context of recommendation systems. The main content of the survey is covered in Section~\ref{sec:scena}, where we conduct separate reviews of the three self-supervised learning paradigms across various recommendation scenarios. Moving forward, Section~\ref{sec:future} delves into open problems and future directions in this field. Lastly, we conclude the survey with Section~\ref{sec:conclusion}. \\\vspace{-0.3in}

%% file: preli.tex
\section{Preliminaries and Definition}
\label{sec:preli}

In this section, we provide a brief introduction to the background relevant to our survey on self-supervised learning for recommendations. We start with an overview of the tasks in recommender systems and then define self-supervised learning, introducing key paradigms: contrastive, reconstructive, and adversarial approaches.
\\\vspace{-0.3in}

\subsection{Recommender Systems}
Recommendation research encompasses diverse tasks including collaborative filtering, sequential recommendation, and multi-behavior recommendation. We provide a general definition applicable across these scenarios. In recommender systems, there are two primary sets: users $\mathcal{U} = \{u_1, ..., u_i, ..., u_{|\mathcal{U}|} \}$ and items $\mathcal{V} = \{v_1, ..., v_j, ..., v_{|\mathcal{V}|} \}$.
An interaction matrix $\mathcal{A} \in \mathbb{R}^{|\mathcal{U}| \times |\mathcal{V}|}$ represents recorded user-item interactions, where $\mathcal{A}_{i, j} = 1$ if user $u_i$ interacted with item $v_j$, and $0$ otherwise. Various recommendation tasks incorporate distinct auxiliary data $\mathcal{X}$ (\eg, knowledge graphs in KG-enhanced recommendation, user relationships in social recommendation).
A recommendation model optimizes a prediction function $f(\cdot)$ that estimates preference scores: $y_{u,v} = f(\mathcal{A}, \mathcal{X}, u, v)$. The score $y_{u, v}$ indicates interaction likelihood, enabling item ranking and recommendation. In Section~\ref{sec:scena}, we will explore the data formulation of $(\mathcal{A}, \mathcal{X})$ under various recommendation scenarios, complementing the demonstration of self-supervised learning in recommendation.
\\\vspace{-0.3in}
 
\subsection{Self-supervised Learning in Recommendation}
Over the past years, deep neural networks have demonstrated outstanding performance with supervised learning in various fields, including computer vision~\cite{voulodimos2018deep}, natural language processing~\cite{deng2018deep} and recommender systems~\cite{zhang2019deep}. However, due to the heavy reliance on labeled data, supervised learning faces challenges when dealing with label sparsity, which is also a common issue in recommendation systems~\cite{SGL, yu2023self}. To address this limitation, self-supervised learning has emerged as a promising approach, leveraging the data itself as labels for learning. In this section, we introduce three fundamental self-supervised learning methodologies: contrastive, reconstructive, and adversarial. \\\vspace{-0.3in}

\subsubsection{\textbf{Contrastive Learning}}

To fully leverage the inherent information within the data itself as supervision signals, contrastive learning has emerged as a prominent self-supervised learning approach~\cite{jaiswal2020survey}. The primary objective of contrastive learning is to maximize the agreement between different views augmented from the data. Formally, in contrastive learning for the recommendation, the objective is to minimize the following loss function~\cite{wu2021self, xie2022self}:
\begin{align}
    \mathop{\min}_{\mathcal{E}_1, \mathcal{E}_2}\ \mathcal{L}_{con}(\mathcal{E}_1 \circ \omega_1(\mathcal{A}, \mathcal{X}),\ \mathcal{E}_2 \circ \omega_2(\mathcal{A}, \mathcal{X})).
\end{align}
Here, $\mathcal{E}_* \circ \omega_*$ represents view creation operations that vary among different methods. Each operation includes data permutation processes, $\omega_1(\cdot)$ and $\omega_2(\cdot)$, which may involve dropping nodes/edges in graphs, along with embedding encoding processes, $\mathcal{E}_1$ and $\mathcal{E}_2$. The goal of minimizing $\mathcal{L}_{con}$ is to derive robust encoding functions that maximize agreement between views, achieved through methods like mutual information maximization or instance discrimination.
\\\vspace{-0.3in}

\subsubsection{\textbf{Reconstructive Learning}}

Reconstructive learning seeks to understand data patterns to learn meaningful representations. It optimizes a deep encoder-decoder model that reconstructs missing or corrupted input data. The encoder, $\mathcal{E}(\cdot)$, creates latent representations from the input, while the decoder, $\mathcal{D}(\cdot)$, reconstructs the original data from the encoder output. The goal is to minimize the discrepancy between reconstructed and original data~\cite{liu2022graph} as follows:
\begin{align}
    \mathop{\min}_{\mathcal{D}, \mathcal{E}} \ \mathcal{L}_{gen}(\mathcal{D} \circ \mathcal{E}(\omega(\mathcal{A}, \mathcal{X})),\  (\mathcal{A}, \mathcal{X})).
\end{align}
Here, $\omega$ represents operations like masking. $\mathcal{D} \circ \mathcal{E}$ denotes the encoding and decoding process for output reconstruction. Recent studies introduce a decoder-only architecture that reconstructs data using a single model (\eg, Transformer~\cite{vaswani2017attention}), commonly applied in sequential recommendation with reconstructive learning~\cite{BERT4Rec}. The loss function $\mathcal{L}_{gen}$ varies by data type, using mean square loss for continuous data and cross-entropy loss for categorical data.
\\\vspace{-0.3in}

\subsubsection{\textbf{Adversarial Learning}} Adversarial learning is a training method used to generate high-quality outputs, using a generator $\mathcal{G}(\cdot)$. What sets adversarial learning apart from reconstructive learning is the inclusion of a discriminator $\Omega(\cdot)$, which determines whether a given sample is real or generated~\cite{gui2021review, jabbar2021survey}. In adversarial learning, the generator aims to enhance the quality of its generated outputs in order to deceive the discriminator. Consequently, the learning objective of adversarial learning can be defined as follows:
\begin{align}
    \underset{\mathcal{G}}{\min}\  \underset{\Omega}{\max}\  \{\mathbb{E}_{x \sim P(\mathcal{A}, \mathcal{X})}[\log\ {\Omega}(x)] + \mathbb{E}_{\hat{x} \sim P(\mathcal{G}(\mathcal{A}, \mathcal{X}))}[\log\ (1 - {\Omega}(\hat{x}))]\}. \label{eq:4}
\end{align}
Here, the variable $x$ represents a real sample from the data distribution, while $\hat{x}$ denotes a synthetic sample generated by the generator $G(\cdot)$. During training, both the generator and discriminator improve through competitive interplay. Ultimately, the generator aims to produce high-quality outputs beneficial for downstream tasks. This approach is commonly used across various domains, including sequential recommendation, to enhance performance.\\\vspace{-0.2in}

\subsubsection{\textbf{Interpretability and Explainability}} In recommender systems, interpretability~\cite{KGIN} and explainability~\cite{ExplainableRec-Survey} have emerged as critical concepts for elucidating the mechanisms of black-box neural models. Interpretability emphasizes the intrinsic transparency of a recommender's structure and decision-making processes, while explainability pertains to the post hoc generation of human-understandable rationales for the model's recommendations. In the context of self-supervised learning, these techniques enhance our understanding of two essential attributes of modern recommenders:  (i) For interpretability, some methods~\cite{zhou2023contrastive} construct structured causal graphs to facilitate an interpretable recommendation process and employ contrastive learning to mitigate exposure bias. Others~\cite{KGIN, DiffMM, DiffKG, DCCF} derive specific module architectures that inherently possess interpretability and utilize self-supervised learning to optimize model parameter training.  (ii) Regarding explainability, prominent approaches~\cite{zhuang2024improving, MMCT, SERMON} leverage diverse contrastive learning techniques to improve the modeling and learning processes of explainable recommendations. Additionally, some methods~\cite{EC4SRec} utilize explanations generated by models for data augmentation in contrastive learning or develop multiple discriminators~\cite{MFGAN} in adversarial learning to elucidate recommendation behaviors. \\\vspace{-0.2in}

\subsubsection{\textbf{Robustness against Noise}} Self-supervised learning, originally developed to identify patterns in large-scale unlabeled datasets, inherently possesses the ability to mitigate the negative effects of noise, which is often unavoidable in such data. By integrating self-supervised learning techniques into recommender systems, we can enhance the robustness of models against both intentional and feature-specific noise present in recommendation data. For instance, in adversarial learning, several methods introduce adversarial perturbations~\cite{yuan2019adversarial} to the original data, thereby training more robust models that can withstand noise. Additionally, in contrastive learning, training models using mutual information lower bounds~\cite{KGCL, HCCF, AdaGCL} with positive-negative discrimination has demonstrated significant advantages in handling noisy data. Some methods~\cite{SimGCL, XSimGCL} also incorporate noise into the augmentation process to further improve training outcomes. Furthermore, in reconstructive learning, the task of reconstructing the original data patterns from noisy inputs inherently equips the model with resilience to noise. Representative methods in this domain include variational auto-encoders~\cite{CVAE} and diffusion-based approaches~\cite{DiffRec, DiffKG}.

%% file: taxon.tex
\section{Taxonomy}
\label{sec:taxonomy}

\begin{figure*}
    \centering
    \includegraphics[width=0.95\textwidth]{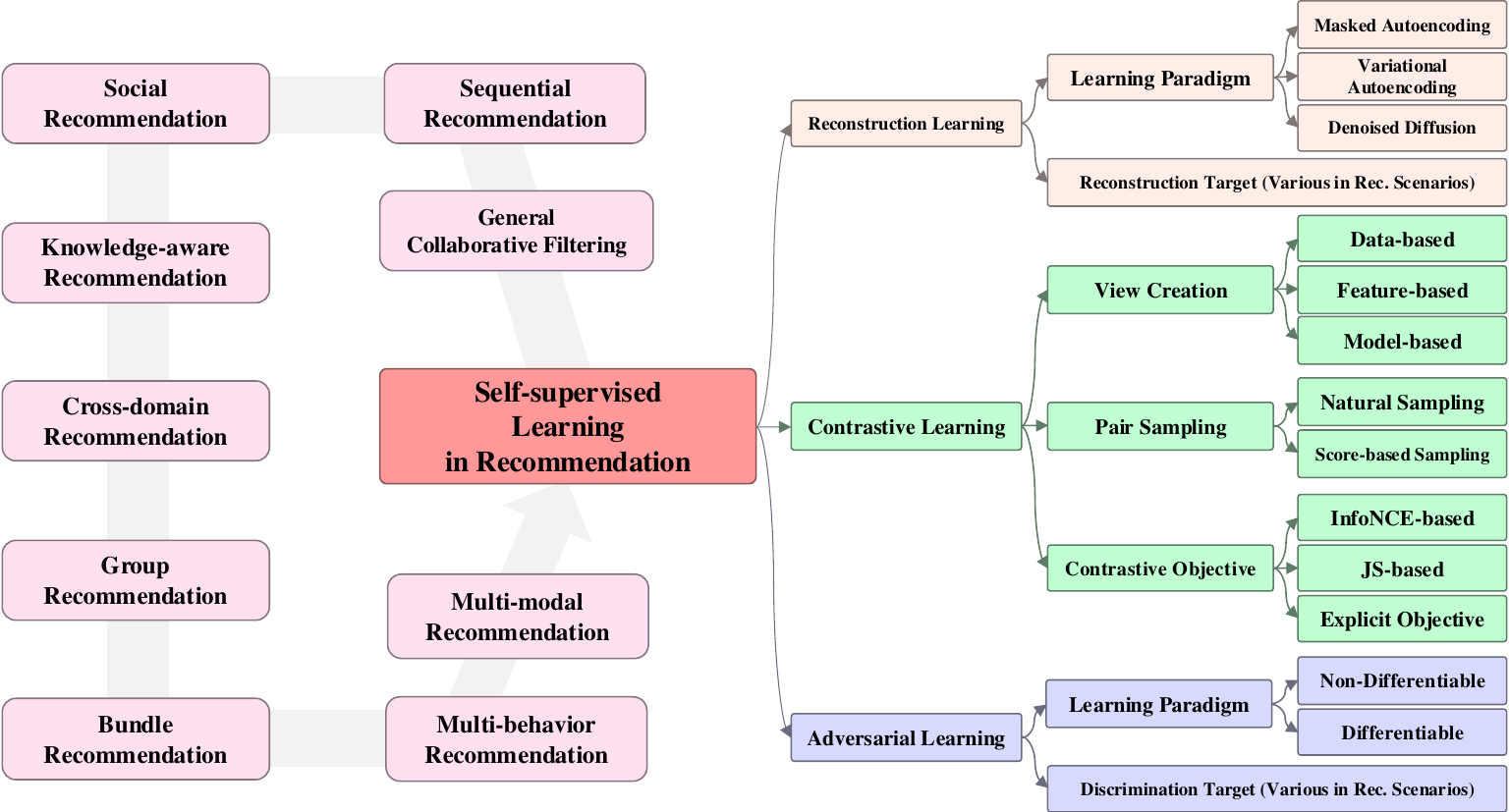}
    \vspace{-0.15in}
    \caption{The proposed taxonomy of self-supervised learning in recommender systems.}
    \label{fig:taxonomy}
    \vspace{-0.25in}
\end{figure*}

In this section, we present our comprehensive taxonomy of self-supervised learning in recommender systems. Different from previous works~\cite{jing2023contrastive, yu2023self} that either focus solely on contrastive learning or categorize SSL into contrastive, generative, predictive and hybrid methods, we categorize SSL into three main learning paradigms (\textit{i.e.}, contrastive, reconstructive and adversarial) and propose detailed taxonomies within each paradigm. Our framework is built upon these three categories, providing comprehensive insights into each approach's methodologies and applications. Figure~\ref{fig:taxonomy} illustrates this organizational structure and highlights the key distinctions between different SSL paradigms. \\\vspace{-0.3in}

\subsection{Contrastive Learning in Recommendation}\label{sec:taxo_cl}
\begin{figure*}[t]
    \centering
    \includegraphics[width=1.0\textwidth]{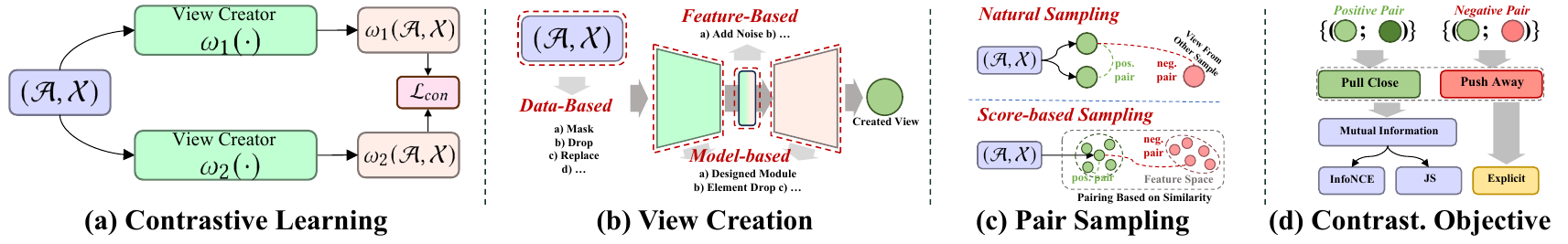}
    \vspace{-0.25in}
    \caption{Taxonomy of Contrastive Learning in Recommender Systems.}
    \vspace{-0.25in}
    \label{fig:cl paradigm}
\end{figure*}

The fundamental principle of contrastive learning (CL) involves maximizing the agreement between different views. Hence, we propose a view-centric taxonomy, as illustrated in Figure~\ref{fig:cl paradigm}, which contains three key components to consider when applying CL: creating the views, pairing the views to maximize agreement, and optimizing the agreement. \\\vspace{-0.3in}

\subsubsection{\textbf{View Creation}} The created view emphasizes various data aspects for the model. It can incorporate global collaborative information to improve the recommender's handling of global relationships~\cite{HCCF} or introduce random noise to enhance model robustness~\cite{SimGCL}. We regard augmentation on input data (\eg, graphs, sequences, input features) \cite{yu2023self, jing2023contrastive} as data-perspective view creation, and hidden feature augmentation during inference as feature-level view creation. Additionally, the model-based contrastive paradigm~\cite{yu2023self} serves as model-level view generation. Thus, we propose a hierarchical taxonomy encompassing view-creation techniques from basic data-level to neural model-level.
\begin{itemize}
    \item \textbf{Data-based view creation.} In the realm of contrastive learning-based recommenders, diverse views are created by augmenting input data. These augmented data points are subsequently processed through deep neural recommenders. The resulting output embeddings from different views are then paired and utilized for contrastive learning. The augmentation methods for the original data vary depending on the recommendation scenario. For instance, graph-based data may employ techniques such as node/edge dropout or the addition of noisy edges~\cite{SGL, SelfCF}, while item sequences may make use of masking, cropping, and replacing~\cite{BERT4Rec, S3-Rec}.
    
    \item \textbf{Feature-based view creation.} In addition to generating views directly from the data, some methods consider conducting augmentation based on the hidden feature encoded during the models' forward process. These representations can include node embedding during graph-based message passing or token vectors in the Transformer, for example. By applying various augmentation techniques or introducing random perturbations to the representations multiple times~\cite{SGL, XSimGCL}, the model's final output can be considered as different views. One commonly used practice is to add random noise to the representations~\cite{ACVAE}.

    \item \textbf{Model-based view creation.} Both data-based and feature-based augmentation are non-adaptive since they are non-parametric. Consequently, various sophisticated methods have emerged to generate different views using neural modules. These views contain specific information based on the model design. For example, intent disentanglement neural modules can capture user intents~\cite{DCCF, wu2023dual}, while hyperparagraph graph modules can capture global relationships~\cite{HCCF}. In contrast, model-based view creation involves learnable parameters within the view generator that are adaptive to the learning objective and optimized during the learning process. \\\vspace{-0.3in}
\end{itemize}

\subsubsection{\textbf{Pair Sampling}} The view creation process generates at least two distinct views for each sample in the data using appropriate view creation methods. The crux of contrastive learning lies in maximally aligning certain views (\ie, pulling them close) while pushing others apart. To achieve this, it's crucial to determine the positive pair of views that should be pulled close and identify other views that form negative pairs to be pushed away. This strategy is known as pair sampling, and it primarily consists of two main pair-sampling approaches in CL-based recommendation: \\\vspace{-0.25in}
\begin{itemize}
    \item \textbf{Natural Sampling.} One common approach to pair sampling is straightforward and non-heuristic, which we refer to as natural sampling. Positive pairs are formed by different views generated from the same data sample, while negative pairs are formed by views from different data samples. In cases where a central view exists for all samples, such as a global view derived from the entire graph, the local-global relationship also forms positive pairs. This approach is widely applied in most contrastive learning recommenders.

    \item \textbf{Score-based Sampling.} Another approach to pair sampling is score-based sampling. In this approach, a module calculates scores of pairs to determine positive or negative pairs. For instance, one relevant score is the distance between two views, which can be used to form pairs~\cite{SHT, DuoRec}. Alternatively, clustering can be applied on views, where positive pairs are those within the same cluster, and negative pairs are those in different clusters~\cite{MIC}. \\\vspace{-0.25in}
\end{itemize}
For an anchor view, once the positive pairs are determined, the remaining views can naturally be considered as negative views, which can be paired with the given view to create negative pairs, allowing for pushing away. Therefore, in the subsequent discussion on different methods of pair sampling, we primarily focus on the construction of positive pairs. \\\vspace{-0.3in}

\subsubsection{\textbf{Contrastive Objective}} The learning objective in contrastive learning is to maximize the mutual information between positive views, which, in turn, leads to improved performance in learning recommender models. Since directly calculating mutual information is not feasible, a tractable lower bound is commonly used as the learning objective in contrastive learning~\cite{belghazi2018mutual}. However, there are also explicit objectives that directly pull positive views closer together. \\\vspace{-0.25in}
\begin{itemize}
    \item \textbf{InfoNCE-based Objective.} InfoNCE~\cite{oord2018representation}, a variant of Noise Contrastive Estimation~\cite{gutmann2010noise}, has gained wide adoption as a learning objective in the field of contrastive learning for recommendation systems. The mathematical formulation of InfoNCE can be expressed as follows:
    \begin{align}
        \mathcal{L} = \mathbb{E}[- \log \frac{exp(f(\omega'_i, \omega''_i))}{\sum_{\forall i, j} exp(f(\omega'_i, \omega''_j))}]
    \end{align}
    Here, $f(\cdot)$ represents a critic function that calculates a score indicating the similarity between two views. The term $f(\omega'_i, \omega''_i)$ corresponds to the score of positive pairs, while the term $\sum \exp(f(\omega'_i, \omega''_j))$ encompasses both the numerator and the scores of all negative pairs. By optimizing this, the $f(\cdot)$ will learn to assign higher values to positive pairs. This process aims to bring the positive pairs closer together and push the negative pairs apart.
    
    \item \textbf{JS-based Objective.} In addition to using InfoNCE estimation for mutual information, the lower bound can also be estimated using the Jensen-Shannon (JS) divergence~\cite{hjelm2018learning, jing2023contrastive}. The derived learning objective is akin to combining InfoNCE with a standard binary cross-entropy loss~\cite{DHCN}, applied to positive pairs and negative pairs:
    \begin{align}
        \mathcal{L} = \mathbb{E}[-\log \sigma(f(\omega'_i, \omega''_i))] - \mathbb{E}[\log (1 - \sigma(f(\omega'_i, \omega''_j)))]
    \end{align}
    Here, $\sigma$ represents the sigmoid function used to normalize the output of the critic function. The main idea behind this optimization is to assign the label $1$ to positive pairs and $0$ to negative pairs, thereby increasing the predicted value for positive pairs and enhancing the similarity between them.
    
    \item \textbf{Explicit Objective.} Both InfoNCE-based and JS-based objectives maximize the estimated lower bound of mutual information to maximize mutual information itself, which is theoretically guaranteed. Additionally, there are explicit objectives, such as minimizing Mean Square Error or maximizing cosine similarity between samples within a positive pair, that directly align positive pairs. These objectives are referred to as explicit objectives. \\\vspace{-0.25in}
\end{itemize}

\subsection{Reconstructive Learning in Recommendation}
\begin{figure*}[t]
    \centering
    \includegraphics[width=1.0\textwidth]{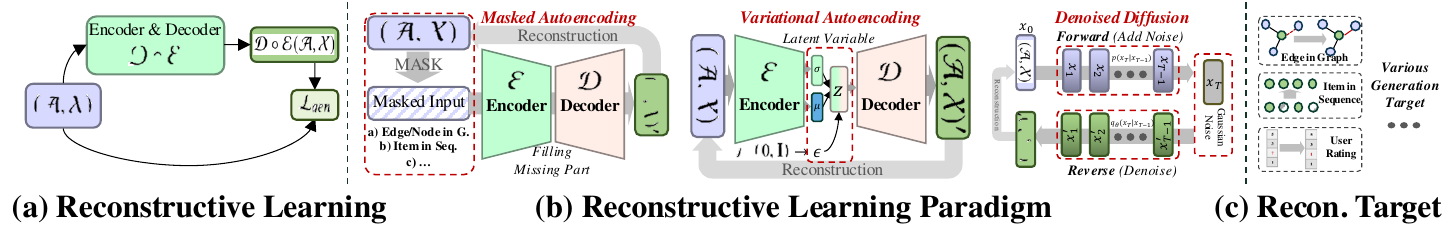}
    \vspace{-0.25in}
    \caption{Taxonomy of Reconstructive Learning in Recommender Systems.}
    \vspace{-0.15in}
    \label{fig:gl paradigm}
\end{figure*}

In reconstructive self-supervised learning (RL), the primary objective is to maximize the likelihood estimation of the real data distribution. This allows the learned meaningful representations to capture the underlying patterns in the data, which can then be utilized for downstream tasks. In our taxonomy (as illustrated in Figure~\ref{fig:gl paradigm}), we consider two aspects to differentiate various recommendation methods with RL: reconstructive learning paradigm and generation target.\\\vspace{-0.25in}

\subsubsection{\textbf{Learning Paradigm}}
In the context of recommendation, recent SSL methods employing reconstructive learning can be classified into three paradigms: Masked Autoencoding, Variational Autoencoding and Denoised Diffusion. \\\vspace{-0.25in}
\begin{itemize}
    \item \textbf{Masked Autoencoding.} In masked autoencoders, the learning process follows a mask-reconstruction approach, where the model reconstructs complete data from partial observations~\cite{he2022masked}. In recommender systems, data can be in item sequences or user-item interaction graphs, each requiring specific masking techniques. For item sequences, certain items are randomly masked and fed into a transformer model to encode representations and reconstruct masked item features~\cite{BERT4Rec}. For graph data, masking occurs at the edge or node level. In GraphMAE~\cite{GraphMAE}, node features are masked for reconstruction, while in MaskGAE~\cite{MaskGAE}, edges are masked. This masking practice is common in reconstructive self-supervised learning recommender models.
    \item \textbf{Variational Autoencoding.} The variational autoencoder~\cite{VAE} is another reconstructive approach that maximizes the likelihood estimation with theoretical guarantees. Typically, it involves mapping the input data to latent factors that follow a normal Gaussian distribution. Subsequently, the model reconstructs the input data based on sampled latent factors. Naturally, during the optimization process of VAE-based models, the learned latent factors are regularized using KL-divergence to enforce them to follow a Gaussian distribution. In self-supervised learning methods employing variational autoencoders, the latent factors may further be regulated through additional approaches, such as enhancing distinguishability through adversarial learning or contrastive learning~\cite{ContrastVAE}.
    \item \textbf{Denoised Diffusion.} Denoising diffusion is a generative model that reconstructs the data samples by reversing a noising process. In the forward process, Gaussian noise is added to the original data in multiple steps, creating a sequence of noisy versions. In the reverse process, the model learns to remove the noise from the noisy versions, recovering the original data step by step. The model is trained to denoise the data at each step, learning to capture minor changes in the complex generation process. The diffusion model has recently been introduced into the recommendation for processing data like graphs~\cite{DiffKG} or user interactions~\cite{DiffRec}.
\end{itemize}

\subsubsection{\textbf{Reconstruction Target}}
What pattern of the data will be considered as the label for reconstruction in reconstructive learning is yet another thing that needs to be considered to bring meaningful auxiliary self-supervised signals. In general, the target can vary for different methods and also in different recommendation scenarios. For instance, in sequential recommendation, the reconstruction target can be the item in the sequence~\cite{BERT4Rec, SVAE}, in order to model the relationships within the item sequence. For recommendation with interaction graph, the target can be the node/edge in the graph~\cite{AutoCF, GFormer}, in order to capture high-level topological relevance in the graph.

\subsection{Adversarial Learning in Recommendation}
\begin{figure*}[t]
    \centering
    \includegraphics[width=0.95\textwidth]{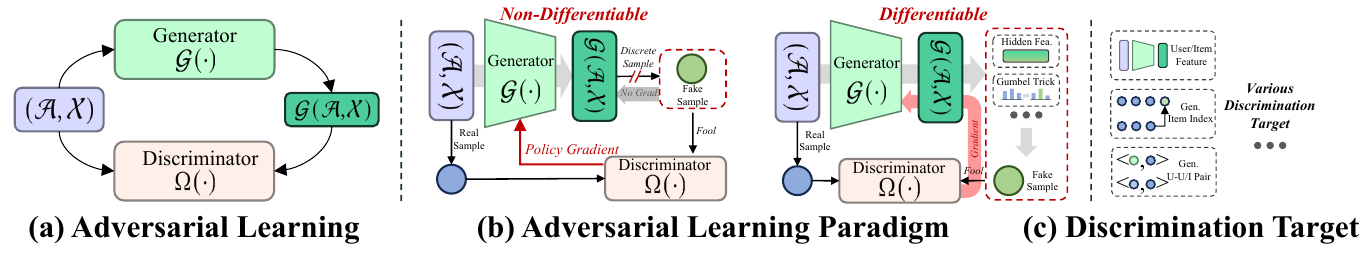}
    \vspace{-0.1in}
    \caption{Taxonomy of Adversarial Learning in Recommender Systems.}
    \vspace{-0.25in}
    \label{fig:al paradigm}
\end{figure*}

In the context of adversarial learning (AL) for recommender systems, the discriminator serves as a critical component that distinguishes between generated fake samples and authentic real samples. Drawing parallels to reconstructive learning, we establish our taxonomy (as depicted in Figure~\ref{fig:al paradigm}) that characterizes AL-based recommendation methods from two key perspectives: the learning paradigm and the discrimination target. \\\vspace{-0.25in}

\subsubsection{\textbf{Learning Paradigm}}
In recommender systems, AL consists of two different paradigms, depending on whether the discriminative loss from the discriminator can be backpropagated to the generator in a differentiable way. \\\vspace{-0.25in}
\begin{itemize}
    \item \textbf{Differentiable AL} The first approach involves objects represented as features in continuous space, where the gradient from the discriminator can naturally backpropagate to the generator for optimization. This method is known as differentiable adversarial learning. Here, the optimization loss aligns directly with the general learning objective in Equation (\ref{eq:4}) using a neural network-based discriminator. The discriminated objects vary depending on the method, such as latent representations from a variational autoencoder's forward process, output features from a Transformer model, or generated user-item interaction matrices. As the quality of the generated content improves, the recommender system's performance is simultaneously enhanced.
    \item \textbf{Non-differentiable AL} Another approach involves discriminating the output of the recommender system, specifically the recommended items. However, the discrete nature of these results complicates back-propagation, making it non-differentiable and preventing the gradient from being directly propagated to the generator~\cite{IRGAN, AOS4Rec}. To address this, reinforcement learning with policy gradients is used~\cite{sutton1999policy}. Here, the generator acts as an agent that predicts the next item based on prior interactions, while the discriminator serves as the reward function, guiding the generator's learning. The rewards emphasize various factors affecting recommendation quality, optimizing to assign higher rewards to true samples compared to generated ones, thus encouraging high-quality recommendations. \\\vspace{-0.2in}
\end{itemize}

\subsubsection{\textbf{Discrimination Target}}
Different recommendation algorithms produce varied inputs for the generator, which are then evaluated by the discriminator. This process enhances the generator's ability to create high-quality content that aligns with the ground truth. Discrimination objectives are tailored to specific recommendation tasks. For example, in sequential recommendation, the discriminator may assess generated next-items~\cite{SSRGAN} or entire sequences. In other areas, it may oversee data quality or optimize the model's hidden features~\cite{su2022cross, DA-CDR}. \\\vspace{-0.3in}

\subsection{Analysis of Self-supervised Learning Techniques}In the realm of self-supervised learning (SSL) for recommender systems, various techniques-namely adversarial, reconstructive, and contrastive learning-offer distinct advantages tailored to specific challenges. Here, we provide an in-depth analysis of the merits of each method and the trends in the development of SSL techniques for recommendation:
\begin{itemize}
    \item \textbf{Merits of Diverse SSL Techniques.} Adversarial learning enhances model robustness by using a discriminator to distinguish between real and generated samples, which is crucial for maintaining data quality. Consequently, this method encourages the generator to produce high-fidelity recommendations, while the integration of reinforcement learning provides adaptive feedback for further refinement. In contrast, reconstructive learning focuses on capturing the underlying data distribution through techniques such as variational autoencoders and masked autoencoders. This approach is effective for modeling complex structures and enables data reconstruction, thereby facilitating augmentation that improves overall model performance. Meanwhile, contrastive learning prioritizes maximizing agreement between different views of the same data, effectively distinguishing positive and negative pairs. As a result, this technique captures nuanced relationships and enhances generalization, particularly in scenarios where labeled data is sparse. In summary, while adversarial and reconstructive techniques improve data quality and representation, contrastive learning excels at extracting meaningful insights from unlabeled datasets. Hence, the choice of techniques usually align with the specific requirements of the task.
    \item \textbf{Evolution in SSL-based Recommendation.} The evolution of self-supervised learning (SSL) techniques in recommendation has generally paralleled advancements in self-supervised methods within deep learning. Prior to 2021, SSL algorithms primarily relied on variational autoencoders~\cite{CVAE, Mult-VAE} and adversarial learning~\cite{SAO, SRecGAN}, significantly enhancing model performance and robustness. Following 2021, two major trends emerged: first, the development of masked autoencoding~\cite{GraphMAE} and reconstructive learning~\cite{BERT4Rec} led to effective applications in sequential recommendation; second, the rise and theoretical refinement of contrastive learning~\cite{SGL, HCCF} resulted in numerous research studies focused on contrastive-based recommendation methods. Recently, the introduction of diffusion models~\cite{DiffRec} has emerged as a new trend in the study of SSL-based recommendation. \\\vspace{-0.3in}
\end{itemize}

%% file: scena.tex
\section{Self-Supervised Learning for Recommendation}
\label{sec:scena}
The study of recommendation algorithms encompasses a variety of scenarios. In this paper, we conducted a survey specifically addressing ten scenarios utilizing self-supervised learning: General Collaborative Filtering, Sequential Recommendation, Social Recommendation, Knowledge-aware Recommendation, Cross-domain Recommendation, Group
Recommendation, Bundle Recommendation, Multi-behavior Recommendation, and Multi-modal Recommendation. \\\vspace{-0.25in}

\subsection{General Collaborative Filtering}

\begin{table*}[!tb]
    \centering
    \small
    \caption{A summary of SSL recommendation methods in \textbf{General Collaborative Filtering} (Part 1). For alphabets in "Category", CL means Contrastive Learning; RL means Reconstructive Learning; AL means Adversarial Learning}
    \vspace{-0.15in}
    \begin{tabular}{c|c|c|c|c|c|c}    
        \toprule
        \multicolumn{7}{c}{\textbf{General Collaborative Filtering (CF)}} \\
        \midrule
        Category & \multicolumn{3}{c|}{Method Information} & \multicolumn{3}{c}{Self-supervised Learning Paradigm} \\
        \midrule
        \multirow{26}{*}{CL} & Method & Year & Venue & View Creation & Pair Sampling & Contrastive Objective \\
        \cmidrule{2-7}
        & Liu \textit{et al.}~\cite{liu2021contrastive}                    & 2021  & arXiv         & Data-based            & Natural                   & InfoNCE-based\\
        & SimpleX~\cite{SimpleX}                & 2021  & CIKM          & Model-based           & Natural                   & Explicit-based\\
        & EGLN~\cite{EGLN}                      & 2021  & SIGIR         & Data \& Model-based   & Natural                   & JS-based\\
        & SGL~\cite{SGL}                        & 2021  & SIGIR         & Data-based            & Natural                   & InfoNCE-based\\
        & BiGI~\cite{BiGI}                      & 2021  & WSDM          & Data \& Model-based   & Natural                   & JS-based\\
        & GDCL~\cite{GDCL}                      & 2022  & DASFAA        & Model-based           & Natural                   & InfoNCE-based\\
        & DirectAU~\cite{DirectAU}              & 2022  & KDD           & Model-based           & Natural                   & Explicit\\
        & SHT~\cite{SHT}                        & 2022  & KDD           & Model-based           & Score-based               & Explicit\\
        & HCCF~\cite{HCCF}                      & 2022  & SIGIR         & Model-based           & Natural                   & InfoNCE-based\\
        & RGCF~\cite{RGCF}                      & 2022  & SIGIR         & Data-based            & Natural                   & InfoNCE-based\\
        & SimGCL~\cite{SimGCL}                  & 2022  & SIGIR         & Feature-based         & Natural                   & InfoNCE-based\\
        & NCL~\cite{NCL}                        & 2022  & WWW           & Model-based           & Natural                   & InfoNCE-based\\
        & CCL~\cite{zhou2023contrastive}        & 2023  & CIKM          & Data-based            & Score-based               & InfoNCE-based\\
        & LightGCL~\cite{LightGCL}              & 2023  & ICLR          & Data-based            & Natural                   & InfoNCE-based\\
        & AdaGCL~\cite{AdaGCL}                  & 2023  & KDD           & Data-based            & Natural                   & InfoNCE-based\\
        & AdvInfoNCE~\cite{AdvInfoNCE}          & 2023  & NeurIPS       & Model-based           & Natural                   & InfoNCE-based\\
        & AdaMCL~\cite{AdaMCL}                  & 2023  & SIGIR         & Data-based            & Natural                   & InfoNCE-based\\
        & CGCL~\cite{CGCL}                      & 2023  & SIGIR         & Model-based           & Natural                   & InfoNCE-based\\
        & DCCF~\cite{DCCF}                      & 2023  & SIGIR         & Model-based           & Natural                   & InfoNCE-based\\
        & uCTRL~\cite{uCTRL}                    & 2023  & SIGIR         & Model-based           & Natural                   & Explicit\\
        & VGCL~\cite{VGCL}                      & 2023  & SIGIR         & Data-based            & Natural \& Score-based    & InfoNCE-based\\
        & XSimGCL~\cite{XSimGCL}                & 2023  & TKDE          & Feature-based         & Natural                   & InfoNCE-based\\
        & RocSE~\cite{RocSE}                    & 2023  & TOIS          & Data-based            & Natural                   & InfoNCE-based\\
        & SelfCF~\cite{SelfCF}                  & 2023  & TORS          & Data \& Feature-based & Natural                   & Explicit\\
        & DENS~\cite{DENS}                      & 2023  & WSDM          & Model-based           & Natural                   & Explicit\\
        & SGCCL~\cite{SGCCL}                    & 2023  & WSDM          & Data-based            & Natural                   & InfoNCE-based\\
        & RecDCL~\cite{RecDCL}                  & 2024  & WWW           & Data \& Model-based   & Natural                   & Explicit\\
        & DivGCL~\cite{DivGCL}                  & 2025  & AAAI          & Model-based           & Natural                   & InfoNCE-based\\
    \bottomrule
    \multicolumn{7}{r}{{\footnotesize Continued on next page}}
    \end{tabular}\label{tab:cfpart1}
    \vspace{-0.2in}
\end{table*}

\begin{table*}[!tb]
    \centering
    \small
    \caption{A summary of SSL recommendation methods in \textbf{General Collaborative Filtering} (Part 2).}
    \vspace{-0.15in}
    \begin{tabular}{c|c|c|c|c|c|c}    
        \toprule
        \multicolumn{7}{c}{\textbf{General Collaborative Filtering (CF)}} \\
        \midrule
        Category & \multicolumn{3}{c|}{Method Information} & \multicolumn{3}{c}{Self-supervised Learning Paradigm} \\
        \midrule
        \multirow{12}{*}{RL} & Method & Venue & Year & \multicolumn{2}{c|}{Reconstructive Learning Paradigm} & Reconstruction Target \\
        \cmidrule{2-7}
        & CVAE~\cite{CVAE}                      & 2017  & KDD           & \multicolumn{2}{c|}{Variational Autoencoding} & User Rating \& Item Content\\
        & Mult-VAE~\cite{Mult-VAE}              & 2018  & WWW           & \multicolumn{2}{c|}{Variational Autoencoding} & User Rating to Items\\
        & MacridVAE~\cite{MacridVAE}            & 2019  & NeurIPS       & \multicolumn{2}{c|}{Variational Autoencoding} & User Rating to Items\\
        & RecVAE~\cite{RecVAE}                  & 2020  & WSDM          & \multicolumn{2}{c|}{Variational Autoencoding} & User Rating to Items\\
        & BiVAE~\cite{BiVAE}                    & 2021  & WSDM          & \multicolumn{2}{c|}{Variational Autoencoding} & User-Item Interactions\\
        & FastVAE~\cite{FastVAE}                & 2022  & WWW           & \multicolumn{2}{c|}{Variational Autoencoding} & User Rating to Items\\
        & MD-CVAE~\cite{MD-CVAE}                & 2022  & WWW           & \multicolumn{2}{c|}{Variational Autoencoding} & User Rating \& Item Content\\
        & SE-VAE~\cite{SE-VAE}                  & 2022  & WWW           & \multicolumn{2}{c|}{Variational Autoencoding} & User Rating to Items\\
        & CaD-VAE~\cite{CaD-VAE}                & 2023  & SIGIR         & \multicolumn{2}{c|}{Variational Autoencoding} & User Rating to Items\\
        & DiffRec~\cite{DiffRec}                & 2023  & SIGIR         & \multicolumn{2}{c|}{Denoising Diffusion}      & User Rating to Items\\
        & GFormer~\cite{GFormer}                & 2023  & SIGIR         & \multicolumn{2}{c|}{Masked Autoencoding}      & Edge in User-Item Graph\\
        & AutoCF~\cite{AutoCF}                  & 2023  & WWW           & \multicolumn{2}{c|}{Masked Autoencoding}      & Edge in User-Item Graph\\\midrule
        \multirow{9}{*}{AL} & Method & Year & Venue & \multicolumn{2}{c|}{Adversarial Learning Paradigm} & Discrimination Target\\
        \cmidrule{2-7}
        & IRGAN~\cite{IRGAN}                    & 2017  & SIGIR        & \multicolumn{2}{c|}{Non-differentiable}    & Item Index\\
        & CFGAN~\cite{CFGAN}                    & 2018  & CIKM         & \multicolumn{2}{c|}{Differentiable}        & User Purchase Vector\\
        & Krishnan \textit{et al.}~\cite{krishnan2018adversarial}                & 2018  & CIKM         & \multicolumn{2}{c|}{Non-differentiable}    & Popular-niche Item Pair\\
        & ABinCF~\cite{ABinCF}                  & 2019  & AAAI         & \multicolumn{2}{c|}{Differentiable}        & Item Index\\
        & AugCF~\cite{AugCF}                    & 2019  & KDD          & \multicolumn{2}{c|}{Differentiable}        & User-Item-Rating Tuple\\
        & ACAE~\cite{yuan2019adversarial}    & 2019  & SIGIR        & \multicolumn{2}{c|}{Differentiable}        & User Purchase Vector\\
        & RAGAN~\cite{RAGAN}                    & 2019  & WWW          & \multicolumn{2}{c|}{Differentiable}        & User Purchase Vector\\
        & Conv-GCF~\cite{Conv-GCF}              & 2020  & CIKM         & \multicolumn{2}{c|}{Differentiable}        & Latent Interaction Map\\
    \bottomrule
    \end{tabular}\label{tab:cfpart2}
    \vspace{-0.2in}
\end{table*}

\subsubsection{\textbf{Task Formulation}} In General Collaborative Filtering (CF), there is no additional observed data $\mathcal{X}$, in which the model solely relies on user-item interactions $\mathcal{A}$ to generate personalized recommendations for uninteracted items. 

\subsubsection{\textbf{Contrastive Learning in CF}} Contrastive learning, as highlighted in Table~\ref{tab:cfpart1}, encompasses various SSL-based methods in collaborative filtering. These methods intelligently leverage user-item interaction data to generate diverse contrast views, facilitating effective learning in the recommendation system.
\begin{itemize}
    \item \textbf{View Creation} in CL-based CF covers diverse paradigms. (i) \textit{Data-based View Creation}. Existing works employ different augmentation techniques to create views of the data for contrastive learning. For example, Liu \textit{et al.}\cite{liu2021contrastive} and methods like SGL\cite{SGL} and SelfCF~\cite{SelfCF} use edge/node dropout to augment the user-item graph. RGCF~\cite{RGCF} incorporates graph structure learning to generate two different graph views, such as a denoised graph and a diversity graph. Similar ideas are also utilized in EGLN~\cite{EGLN}. AdaMCL~\cite{AdaMCL} and SGCCL~\cite{SGCCL} create user-user and item-item graphs based on the data and then encode views for users and items. LightGCL~\cite{LightGCL} created another user-item graph based on SVD decomposition that captures global relationship modeling. Recently, RecDCL~\cite{RecDCL} interpolates historical and current embeddings to obtain augmented views for batch contrastive learning (ii) \textit{Feature-based View Creation}. In collaborative filtering, various feature-based augmentations involve adding noise to node features during model inference. Methods such as SimGCL~\cite{SimGCL} and XSimGCL~\cite{XSimGCL} introduce controllable random noise to node embeddings during message passing, thereby creating noised views of nodes. Another approach, RocSE~\cite{RocSE}, employs embedding perturbation techniques inspired by adversarial attacks to generate different embedding views. (iii) \textit{Model-based View Creation}. Different methods also employ various neural modules to create views. On one hand, SimpleX~\cite{SimpleX}, DirectAU~\cite{DirectAU}, RecDCL-FCL~\cite{RecDCL}, CGCL~\cite{CGCL}, AdvInfoNCE~\cite{AdvInfoNCE}, uCTRL~\cite{uCTRL} and XSimGCL~\cite{XSimGCL} directly utilize the encoded user/item embeddings from models as views. On the other hand, some methods incorporate carefully designed neural modules to encode views from different perspectives. For example, EGLN~\cite{EGLN} and BiGI~\cite{BiGI} focus on global representation fusion. GDCL~\cite{GDCL} approximates diffusion models on graphs. HCCF~\cite{HCCF} and SHT~\cite{SHT} employ hypergraph global encoding. DCCF~\cite{DCCF} explores intent disentanglement. NCL~\cite{NCL} and CGCL~\cite{CGCL} incorporate diverse neighbor view discovery. VGCL~\cite{VGCL} utilizes variational latent space sampling. DENS~\cite{DENS} incorporates item factor encoding. AdaGCL~\cite{AdaGCL} designed a graph model and a denoising model to create adaptive contrastive views. 
    \item \textbf{Pair Sampling} in this scenario includes both natural sampling and also score-based sampling. (i) \textit{Natural Sampling}. Most of the works in collaborative filtering follow natural sampling of positive samples for the anchor views to create positive pairs. There are three common situations for natural sampling in CF. Firstly, when the model encodes multiple views for one user/item, any two of these views can form a pair~\cite{SGL, LightGCL, SimGCL, HCCF, DCCF}. Secondly, the views from the interacted user and item can form a pair~\cite{DirectAU, uCTRL, AdvInfoNCE}. Lastly, by considering the relevant/irrelevant factor relationship with user-item interaction, pairs of views can be naturally formed~\cite{DENS}. (ii) \textit{Score-based Sampling}. SHT calculates edge solidity scores and samples pairs of edges for learning. VGCL clusters users/items based on embedding similarity and pairs users from the same cluster.
    \item \textbf{Contrastive Objective} utilizes InfoNCE-based or JS-based objectives to optimize mutual information. Notably, AdvInfoNCE improves the weighting of negative pairs in InfoNCE by incorporating hardness scores learned through a min-max game. Some methods explicitly design loss functions to enhance the similarity between positive pairs. This can be achieved through cosine similarity optimization~\cite{SimpleX, SelfCF, DENS}, alignment/uniformity regularization~\cite{DirectAU, uCTRL, RecDCL}, or by explicitly regulating the score difference between pairs~\cite{SHT}.
\end{itemize}

\subsubsection{\textbf{Reconstructive Learning in CF}} In collaborative filtering, certain self-supervised learning (SSL) methods leverage generative learning techniques (as indicated in Table~\ref{tab:cfpart2}) to reconstruct user-item interactions. This approach allows them to obtain self-supervised signals that aid in the learning process of the model.
\begin{itemize}
    \item \textbf{Reconstructive Learning Paradigm} in self-supervised learning CF has evolved from variational autoencoding to masked autoencoding and denoised diffusion. (i) \textit{Variational Autoencoding}. Different VAE-based methods in self-supervised learning CF adopt various perspectives to improve the model. For instance, CVAE~\cite{CVAE} and MD-CVAE~\cite{MD-CVAE} incorporate item content information to regulate the collaborative latent variable modeling. RecVAE~\cite{RecVAE} follows the general paradigm in Mult-VAE~\cite{Mult-VAE} but introduces a novel composite prior distribution and a new approach to setting hyperparameters. MacridVAE~\cite{MacridVAE} and CaD-VAE~\cite{CaD-VAE} focus on modeling the latent factor behind interactions to capture the latent variable distribution. BiVAE~\cite{BiVAE} extends the VAE paradigm to symmetrically generate user-item interactions. FastVAE~\cite{FastVAE} enhances model efficiency with inverted multi-index. SE-VAE~\cite{SE-VAE} extends latent variable modeling with multi-experts and stochastic expert selection. (ii) \textit{Masked Autoencoding}. Both AutoCF~\cite{AutoCF} and GFormer~\cite{GFormer} adopt the mask-reconstruction paradigm, where they automatically identify valuable edges (user-item interactions) in the graph and reconstruct them using the learned user/item representations. (iii) \textit{Denoised Diffusion}. DiffRec~\cite{DiffRec} has introduced the diffusion model into collaborative filtering. This method leverages diffusion to denoise perturbed user ratings or latent variables, which helps mitigate the high resource costs associated with large-scale item prediction.
    \item \textbf{Reconstruction Target}. Both VAE-based methods~\cite{Mult-VAE, MacridVAE, RecVAE, FastVAE, SE-VAE, CaD-VAE} and diffusion-based methods~\cite{DiffRec} share a similar auto-encoding architecture. In these methods, the user ratings to items, typically represented as vectors, serve as the reconstruction target for the model. However, some VAE-based methods like CVAE and MD-CAE also generate item content to achieve hybrid generation. BiVAE adopts a symmetric approach by modeling the user and item latent variables, with user-item interactions as the reconstruction target. On the other hand, in graph-based methods~\cite{AutoCF, GFormer}, valuable edges in the graph are masked and then reconstructed. \\\vspace{-0.3in}
\end{itemize}

\subsubsection{\textbf{Adversarial Learning in CF}} Collaborative filtering methods also leverage adversarial learning (as shown in Table~\ref{tab:cfpart2}) to improve recommendation indices, user ratings, or even perform data augmentation. These methods optimize the model with discriminator and enhance the overall performance of collaborative filtering systems. \\\vspace{-0.2in}
\begin{itemize}
    \item \textbf{Adversarial Learning Paradigm} in collaborative filtering encompasses both non-differentiable and differentiable training approaches. IRGAN~\cite{IRGAN} introduced the idea of GANs into recommendation systems, sampling item indices from the generator for discrimination. However, the discrete nature of these item indices prevents the gradient from the discriminator from propagating to the generator for optimization. To overcome this issue, IRGAN employs policy gradient techniques, which are also adopted by Krishnan \textit{et al.}~\cite{krishnan2018adversarial}. In the context of differentiable adversarial training, several methods such as CFGAN~\cite{CFGAN}, RAGAN~\cite{RAGAN}, and Conv-GCF~\cite{Conv-GCF} feed the generated continuous real-valued variables into the discriminator. On the other hand, ABinCF~\cite{ABinCF} and AugCF~\cite{AugCF} use the Gumbel trick to approximate sampling, enabling end-to-end adversarial training.
    \item \textbf{Discrimination Target} varies across different methods in collaborative filtering. IRGAN samples the indices of items relevant to a given user for discrimination and ABinCF follows the similar paradigm but makes the parameters of generator binary codes for efficiency. Additionally, CFGAN improves IRGAN by enabling gradient back-propagation through discriminating the user purchase vector generated by the generator. RAGAN extends this further by incorporating low ratings for items in the user purchase vector to address biased predictions. Krishnan \textit{et al.}~\cite{krishnan2018adversarial} leverage popular-niche item pairs for discrimination to improve long-tail recommendation. AugCF treats each user-item-rating tuple as a sample for the discriminator, while Conv-GCF encodes the latent interaction map of a user-item pair using a convolutional neural network for further discrimination. These methods employ different approaches to enhance recommendation performance in collaborative filtering. \\\vspace{-0.3in}
\end{itemize}

\subsubsection{\textbf{In-depth Analysis of SSL-based Collaborative Filtering}} The three main SSL techniques-contrastive learning, reconstructive learning, and adversarial learning-offer distinct advantages and challenges in CF. Contrastive learning enhances model robustness by leveraging diverse views from user-item interactions~\cite{SGL, SelfCF}, but its effectiveness can be hindered by the quality of constructed views. Reconstructive learning, through methods like VAEs and MAEs, excels in reconstructing interactions for richer representations~\cite{RecVAE, AutoCF}, though it may face scalability issues in large datasets. Adversarial learning improves recommendation accuracy by using a discriminator to optimize outputs~\cite{IRGAN, CFGAN}, yet its non-differentiable nature can complicate training for CF. Ultimately, the choice of technique should align with the specific needs of the collaborative filtering task, considering factors like data availability and computational resources.

\subsection{Sequential Recommendation}

\begin{table*}[!tb]
    \centering
    \small
    \caption{A summary of SSL recommendation methods in \textbf{Sequential Recommendation} (Part 1).}
    \vspace{-0.15in}
    \begin{tabular}{c|c|c|c|c|c|c}    
        \toprule
        \multicolumn{7}{c}{\textbf{Sequential Recommendation (SeqRec)}} \\
        \midrule
        Category & \multicolumn{3}{c|}{Method Information} & \multicolumn{3}{c}{Self-supervised Learning Paradigm} \\
        \midrule
        \multirow{33}{*}{CL} & Method & Year & Venue & View Creation & Pair Sampling & Contrastive Objective \\
        \cmidrule{2-7}
        & S3-Rec~\cite{S3-Rec}                  & 2020  & CIKM          & Data-based            & Natural                   & InfoNCE-based\\
        & Ma \textit{et al.}~\cite{ma2020disentangled}   & 2020  & KDD           & Data \& Model-based   & Natural                   & InfoNCE-based\\
        & DHCN~\cite{DHCN}                      & 2021  & AAAI          & Model-based           & Natural                   & JS-based\\
        & CoSeRec~\cite{CoSeRec}                & 2021  & ArXiv         & Data-based            & Natural                   & InfoNCE-based\\
        & CCL~\cite{CCL}                        & 2021  & CIKM          & Data-based            & Natural                   & InfoNCE-based\\
        & COTREC~\cite{COTREC}                  & 2021  & CIKM          & Model-based           & Score-based               & InfoNCE-based\\
        & H2SeqRec~\cite{H2SeqRec}              & 2021  & CIKM          & Data-based            & Natural \& Score-based    & InfoNCE-based\\
        & CLUE~\cite{CLUE}                      & 2021  & ICDM          & Data \& Model-based   & Natural                   & Explicit\\
        & MMInfoRec~\cite{MMInfoRec}            & 2021  & ICDM          & Model-based           & Natural                   & InfoNCE-based\\
        & SSI~\cite{SSI}                        & 2021  & IJCAI         & Data-based            & Natural                   & InfoNCE-based\\
        & ACVAE~\cite{ACVAE}                    & 2021  & WWW           & Feature-based         & Natural                   & JS-based\\
        & CBiT~\cite{CBiT}                      & 2022  & CIKM          & Data-based            & Natural                   & InfoNCE-based\\
        & ContrastVAE~\cite{ContrastVAE}        & 2022  & CIKM          & Model-based           & Natural                   & InfoNCE-based\\
        & EC4SRec~\cite{EC4SRec}                & 2022  & CIKM          & Data-based            & Natural                   & InfoNCE-based\\
        & MCLSR~\cite{MCLSR}                    & 2022  & CIKM          & Model-based           & Natural                   & InfoNCE-based\\
        & MIC~\cite{MIC}                        & 2022  & CIKM          & Data \& Feature-based & Score-based               & InfoNCE-based\\
        & TCPSRec~\cite{TCPSRec}                & 2022  & CIKM          & Data-based            & Natural                   & InfoNCE-based\\
        & CL4SRec~\cite{CL4SRec}                & 2022  & ICDE          & Data-based            & Natural                   & InfoNCE-based\\
        & DCAN-PSSL~\cite{DCAN-PSSL}            & 2022  & ICDE          & Model-based           & Natural                   & Explicit\\
        & MISS~\cite{MISS}                      & 2022  & ICDE          & Model-based           & Score-based               & InfoNCE-based\\
        & GCL4SR~\cite{GCL4SR}                  & 2022  & IJCAI         & Data \& Model-based   & Natural                   & InfoNCE-based\\
        & DCN~\cite{DCN}                        & 2022  & SIGIR         & Data-based            & Natural                   & Explicit\\
        & DuoRec~\cite{DuoRec}                  & 2022  & WSDM          & Model-based           & Score-based               & InfoNCE-based\\
        & ICLRec~\cite{ICLRec}                  & 2022  & WWW           & Data \& Model-based   & Natural                   & InfoNCE-based\\
        & TiCoSeRec~\cite{TiCoSeRec}            & 2023  & AAAI          & Data-based            & Natural                   & InfoNCE-based\\
        & ECGAN-Rec~\cite{ECGAN-Rec}            & 2023  & IPM           & Data \& Model-based   & Natural                   & Explicit\\
        & FDSA\_CL~\cite{FDSA_CL}               & 2023  & TKDE          & Model-based           & Natural                   & InfoNCE-based\\
        & ContraRec~\cite{ContraRec}            & 2023  & TOIS          & Data-based            & Natural                   & InfoNCE-based\\
        & MrTransformer~\cite{MrTransformer}    & 2023  & TOIS          & Model-based           & Score-based               & Explicit\\
        & IOCRec~\cite{IOCRec}                  & 2023  & WSDM          & Data \& Model-based   & Natural                   & InfoNCE-based\\
        & DCRec~\cite{DCRec}                    & 2023  & WWW           & Model-based           & Natural                   & InfoNCE-based\\
        & Meta-SGCL~\cite{Meta-SGCL}            & 2024  & ICDE          & Model-based           & Natural                   & InfoNCE-based\\
        & Liu \textit{et al.}~\cite{SA-GNN}     & 2024  & SIGIR         & Model-based           & Natural                   & Explicit\\
        & ICSRec~\cite{ICSRec}                  & 2024  & WSDM          & Model-based           & Natural                   & InfoNCE-based\\
        & LS4SRec~\cite{LS4SRec}                & 2025  & TOIS          & Model-based           & Natural                   & Explicit\\
    \bottomrule
    \multicolumn{7}{r}{{\footnotesize Continued on next page}}
    \end{tabular}\label{tab:seqpart1}
    \vspace{-0.2in}
\end{table*}

\begin{table*}[!tb]
    \centering
    \small
    \caption{A summary of SSL recommendation methods in \textbf{Sequential Recommendation} (Part 2).}
    \vspace{-0.15in}
    \begin{tabular}{c|c|c|c|c|c|c}    
        \toprule
        \multicolumn{7}{c}{\textbf{Sequential Recommendation (SeqRec)}} \\
         \midrule
        Category & \multicolumn{3}{c|}{Method Information} & \multicolumn{3}{c}{Self-supervised Learning Paradigm} \\
        \midrule
        \multirow{14}{*}{RL} & Method & Venue & Year & \multicolumn{2}{c|}{Reconstructive Learning Paradigm} & Reconstruction Target \\
        \cmidrule{2-7}
        & BERT4Rec~\cite{BERT4Rec}              & 2019  & CIKM          & \multicolumn{2}{c|}{Masked Autoencoding}      & Item in Sequence\\
        & SVAE~\cite{SVAE}                      & 2019  & CIKM          & \multicolumn{2}{c|}{Variational Autoencoding} & Item in Sequence\\
        & PTUM~\cite{PTUM}                      & 2020  & EMNLP         & \multicolumn{2}{c|}{Masked Autoencoding}      & Behavior in Sequence\\
        & PeterRec~\cite{PeterRec}              & 2020  & SIGIR         & \multicolumn{2}{c|}{Masked Autoencoding}      & Item in Sequence\\
        & U-BERT~\cite{U-BERT}                  & 2021  & AAAI          & \multicolumn{2}{c|}{Masked Autoencoding}      & Words in Review\\
        & BiCAT~\cite{BiCAT}                    & 2021  & arXiv         & \multicolumn{2}{c|}{Masked Autoencoding}      & Item in Sequence\\
        & ShopperBERT~\cite{ShopperBERT}        & 2021  & arXiV         & \multicolumn{2}{c|}{Masked Autoencoding}      & Item in Sequence\\
        & UPRec~\cite{UPRec}                    & 2021  & arXiv         & \multicolumn{2}{c|}{Masked Autoencoding}      & Item in Sequence\\
        & VSAN~\cite{VSAN}                      & 2021  & ICDE          & \multicolumn{2}{c|}{Variational Autoencoding} & Item in Sequence\\
        & ASReP~\cite{ASReP}                    & 2021  & SIGIR         & \multicolumn{2}{c|}{Masked Autoencoding}      & Item in Sequence\\
        & CBiT~\cite{CBiT}                      & 2022  & CIKM          & \multicolumn{2}{c|}{Masked Autoencoding}      & Item in Sequence\\
        & ContrastVAE~\cite{ContrastVAE}        & 2022  & CIKM          & \multicolumn{2}{c|}{Variational Autoencoding} & Item in Sequence\\
        & DiffuASR~\cite{DiffuASR}              & 2023  & CIKM          & \multicolumn{2}{c|}{Denoised Diffusion}       & Augmented Item Sequence\\
        & Diff4Rec~\cite{Diff4Rec}              & 2023  & MM            & \multicolumn{2}{c|}{Denoised Diffusion}       & User-Item Interaction\\
        & MAERec~\cite{MAERec}                  & 2023  & SIGIR         & \multicolumn{2}{c|}{Masked Autoencoding}      & Edge in Graph\\
        \midrule
        \multirow{8}{*}{AL} & Method & Year & Venue & \multicolumn{2}{c|}{Adversarial Learning Paradigm} & Discrimination Target\\
        \cmidrule{2-7}
        & AOS4Rec~\cite{AOS4Rec}                & 2020  & IJCAI         & \multicolumn{2}{c|}{Non-differentiable} & Item\\
        & MFGAN~\cite{MFGAN}                    & 2020  & SIGIR         & \multicolumn{2}{c|}{Non-differentiable} & Feature of Sequence\\
        & SAO~\cite{SAO}                        & 2020  & SIGIR         & \multicolumn{2}{c|}{Differentiable} & Feature of Sequence\\
        & SRecGAN~\cite{SRecGAN}                & 2021  & DASFAA        & \multicolumn{2}{c|}{Differentiable} & Ranking Score\\
        & SSRGAN~\cite{SSRGAN}                  & 2021  & DASFAA        & \multicolumn{2}{c|}{Differentiable} & Item\\
        & ACVAE~\cite{ACVAE}                    & 2021  & WWW           & \multicolumn{2}{c|}{Differentiable} & Feature of Sequence\\
        & ECGAN-Rec~\cite{ECGAN-Rec}            & 2023  & IPM           & \multicolumn{2}{c|}{Differentiable} & Feature of Sequence\\
    \bottomrule
    \end{tabular}\label{tab:seqpart2}
    \vspace{-0.1in}
\end{table*}

\subsubsection{\textbf{Task Formulation}}
In sequential recommendation (SeqRec), user-item interactions are recorded with timestamps, establishing a temporal order. Each user has a specific temporal sequence of engaged items denoted as $s_{u} = (v_1, v_2, ..., v_T)$, where $T$ represents the sequence length. The objective is to predict the next item ($v_{T+1}$) based on the past item sequence. In anonymous user scenarios, only item sequences are available, known as session recommendation.

\subsubsection{\textbf{Contrastive Learning in SeqRec}} Contrastive learning sequential recommendation comprises the majority of self-supervised SeqRec methods (as indicated in Table~\ref{tab:seqpart1}). These methods leverage the sequential data to generate diverse views through techniques such as augmentation or neural modules.
\begin{itemize}
    \item \textbf{View Creation} in SeqRec includes data-based, feature-based, and model-based methods. (i) \textit{Data-based View Creation.} Given that the input data is in a sequential format, numerous methods employ sequence augmentation techniques to generate diverse views of the input item sequence. In general, early works~\cite{S3-Rec, CL4SRec, CoSeRec, ShopperBERT, H2SeqRec} propose non-heuristic and random augmentation methods including Crop, Mask, Reorder, Shuffle, Substitute, and Insert to obtain different input sequences as different views. TiCoSeRec~\cite{TiCoSeRec} goes a step further by incorporating temporal information and introducing five operators (\eg, Ti-Crop and Ti-Reorder). These operators are designed to transform the input sequence into a uniform representation while taking into account the variations in time intervals. In EC4SRec~\cite{EC4SRec}, importance scores are calculated for each item in a sequence, which is then utilized to guide heuristic and explanation-driven operations for generating informative views. In addition to these methods, several data-based approaches consider constructing additional data resources such as item-item transition graphs~\cite{GCL4SR, DCRec}, item co-interaction~\cite{DCRec}, and user sequences for each item~\cite{DCN} to obtain diverse viewpoints. (ii) \textit{Feature-based View Creation.} Feature-based methods employ various augmentation on the encoded features for view creation. For instance, ACVAE~\cite{ACVAE} utilizes variational auto-encoders to encode latent features of item sequences. It then shuffles these latent features to generate different negative views of the sequence. Additionally, MIC~\cite{MIC} applies random dropout on the encoded features, resulting in two distinct views of user/item representation for contrastive learning. (iii) \textit{Model-based View Creation.} Model-based view creation in SeqRec encompasses two distinct approaches. The first approach involves constructing specific neural modules to encode views of interest, such as intent-aware representation~\cite{ma2020disentangled, ICLRec, ICSRec}, hypergraph representation~\cite{DHCN}, item attributes representation~\cite{FDSA_CL}, variational Transformer~\cite{Meta-SGCL}, long-short term representations~\cite{SA-GNN} and utilizing graph neural networks and to encode embeddings from auxiliary graphs~\cite{MCLSR, DCRec, DCN}. The second approach involves applying techniques such as random dropout~\cite{CLUE, MMInfoRec, ContrastVAE, DuoRec} on the model parameters to generate different outputs as distinct views.
    \item \textbf{Pair Sampling} in SeqRec has both natural sampling and score-based heuristic sampling. (i) \textit{Natural Sampling}. In most cases in SeqRec, natural sampling methods~\cite{S3-Rec, CoSeRec, CCL, CL4SRec, CBiT, EC4SRec, GCL4SR} naturally considering the generated views from the same item sequences as positive views and otherwise negative views. In addition, there are other natural pairing relations utilized in SeqRec. For instance, methods like SSI~\cite{SSI}, TCPSRec~\cite{TCPSRec}, and DCRec~\cite{DCRec} construct positive pairs using various combinations, such as pairing a single item or a sub-sequence with the whole sequence it belongs to. Besides, ICSRec~\cite{ICSRec} uses clustering to generate intent prototypes and aligns the intent view with related prototypes for fine-grained intent CL. (ii) \textit{Score-based Sampling.} Compared to natural sampling, score-based sampling utilizes necessary calculate to determine the positive pair. COTREC~\cite{COTREC} combines the last-clicked item with the predicted next item to form a positive pair. H2SeqRec~\cite{H2SeqRec}, DuoRec~\cite{DuoRec}, and MrTransformer~\cite{MrTransformer} employ sequence overlap as a metric to determine positive pairs. MIC~\cite{MIC} first applies k-means clustering to group samples, and then samples within the same group form positive pairs. Similarly, MISS~\cite{MISS} employs item embedding distance as a metric to sample positive pairs with varying probabilities.
    \item \textbf{Contrastive Objective} in SeqRec typically adopts an InfoNCE-based objective to maximize the mutual information between positive pairs. However, in methods such as DHCN~\cite{DHCN} and ACVAE~\cite{ACVAE}, a JS-based objective is utilized for optimization. Furthermore, several algorithms incorporate explicit loss functions to contrast data samples. The fundamental concept behind these methods is to encourage the proximity of samples within positive pairs. CLUE~\cite{CLUE} leverages cosine similarity to minimize the distance between embeddings within a positive pair. DCAN-PSSL~\cite{DCAN-PSSL} and ECGAN-Rec~\cite{ECGAN-Rec} aim to minimize the estimated probability distribution of the next item. DCAN-PSSL uses KL/JS-divergence minimization, while ECGAN-Rec minimizes the numerical differences between the distributions. DCN~\cite{DCN} and MrTransformer~\cite{MrTransformer} both minimize the squared error between two embeddings within a positive pair for optimization purposes. Liu \textit{et al.}~\cite{SA-GNN} explicitly align the pairwise likelihood difference scores of two edges' long- and short-term node embeddings for denoising.
\end{itemize}

\subsubsection{\textbf{Reconstructive Learning in SeqRec}} Generative learning plays a significant role in sequential recommendation (as demonstrated in Table~\ref{tab:seqpart2}). These methods primarily focus on generating sequence item data or user-item graphs to provide auxiliary signals, effectively enhancing the recommendation process.
\begin{itemize}
    \item \textbf{Reconstructive Learning Paradigm} in SeqRec predominantly follows the approach of mask-autoencoding. For instance, BERT4Rec~\cite{BERT4Rec} initially attempts to randomly mask an item within the item sequence. Then, it proceeds with the encoding process using a transformer and aims to reconstruct the masked item. Given that using transformers for modeling item sequence data~\cite{SASRec} in SeqRec is a prevalent choice, the natural approach of mask-autoencoding with transformers has become a primary form in subsequent works~\cite{PTUM, PeterRec, BiCAT, ShopperBERT, UPRec, ASReP, CBiT}. Furthermore, MAERec~\cite{MAERec} adopts the concept of mask-autoencoding from graph data~\cite{GraphMAE, MaskGAE} to enhance the representation learning of items. Another notable approach in generative modeling is seen in SVAE~\cite{SVAE}, which leverages variational auto-encoding as a backbone. This method utilizes recurrent neural network to encode sequential records into latent variables that follow the normal distribution and subsequently generate the next-item with a re-parametrization trick~\cite{VAE}. This line of work is further extended by methods like VSAN~\cite{VSAN} and ContrastVAE~\cite{ContrastVAE} by incorporating self-attention modules or contrastive learning on latent variables. Recently, denoised diffusion has emerged as a promising paradigm in sequential recommendation (SeqRec) due to its strong generation capabilities. Both DiffuASR~\cite{DiffuASR} and Diff4Rec~\cite{Diff4Rec} leverage diffusion generation to create high-quality data, which in turn enhances the training of recommendation models through data augmentation.
    \item \textbf{Reconstruction Target} in SeqRec with generative learning exhibits a diverse range of approaches. In general, several works~\cite{SVAE, PeterRec, BiCAT, ShopperBERT, UPRec, VSAN, ASReP, CBiT, ContrastVAE} follow the initial step of BERT4Rec, which involves masking and reconstructing the items within the sequence. PTUM considers the behavior sequence of users as the target for generation, aiming to effectively model user behavior patterns for recommendation purposes. On the other hand, U-BERT focuses on generating user reviews for items to capture user behaviors and enhance recommendations, particularly in content-insufficient domains. Recently, MAERec has incorporated the concept of mask-autoencoding from the graph domain. It utilizes this approach to generate paths in the item transition graph, effectively pre-training item embeddings for improved sequential recommendation performance. For diffusion-based methods, DiffuASR generates augmented item sequences to enrich sparse data for model training, while Diff4Rec pre-trains a diffusion model by corrupting and reconstructing user-item interactions in the latent space, which are then used to produce diverse augmentations for sparse user-item interactions.
\end{itemize}

\subsubsection{\textbf{Adversarial Learning in SeqRec}} Adversarial learning in SeqRec (as shown in Table~\ref{tab:seqpart2}) employs various strategies, leveraging encoded features, ranking scores, and historical data to train discriminators in distinguishing between real and generated item sequences, while optimizing generators to produce realistic recommendations.
\begin{itemize}
    \item \textbf{Adversarial Learning Paradigm} in SeqRec encompasses differential and non-differential methods. Differential methods, such as SAO~\cite{SAO}, SRecGAN~\cite{SRecGAN}, ACVAE~\cite{ACVAE}, and ECGAN-Rec~\cite{ECGAN-Rec}, utilize continuous operations for gradient propagation and avoid discrete sampling. SSRGAN~\cite{SSRGAN} employs Gumble-Softmax for differentiable sampling to overcome gradient blocking. In non-differential methods, reinforcement learning optimizes the generator's parameters after generating items through non-differentiable sampling and feeding them to the discriminator. AOS4Rec~\cite{AOS4Rec} employs the WGAN~\cite{WGAN} concept for simultaneous optimization and the actor-critic algorithm~\cite{Actor-critic} for stable training, while MFGAN~\cite{MFGAN} uses policy gradients to improve generator predictions using the discriminator's score as a reward.
    
    \item \textbf{Discrimination Target} exhibits a wide range of diversity within SeqRec when employing adversarial learning. Various approaches, such as MFGAN, SAO, ACVAE, and ECGAN-Rec, employ encoded features derived from the item sequence, which are then fed into the discriminator. The discriminator's role is to determine whether the item sequence is real or generated by the generator. Besides, the discrimination target in SRecGAN is the ranking score of items, where the score difference between positive and negative items generated by the generator deceives the discriminator. The discriminator is also optimized using ground truth data with an ideal score difference, aiming to force the generator to assign higher values to positive pairs. In AOS4Rec, the discriminator calculates the score by considering the history item along with the sampled next-item. It assigns a higher value to the real item sequence. A similar concept is demonstrated in SSRGAN, but it utilizes differential sampling to circumvent the involvement of reinforcement learning.
\end{itemize}

\subsubsection{\textbf{In-depth Analysis of SSL-based Sequential Recommendation}} SSL techniques in sequential recommendation each exhibit unique strengths and limitations tied to the temporal and contextual nature of sequential data. Contrastive learning methods, such as CL4SRec~\cite{CL4SRec} and CoSeRec~\cite{CoSeRec}, excel at capturing nuanced sequential patterns through view augmentations and pairwise alignment, enhancing robustness against noise. However, their performance relies on the design of augmentation strategies and pair sampling, which may inadvertently distort temporal dependencies if not carefully regularized. Reconstructive approaches, exemplified by BERT4Rec~\cite{BERT4Rec} and DiffuASR~\cite{DiffuASR}, leverage reconstruction objectives to model long-term dependencies via transformers or diffusion models, effectively addressing data sparsity through synthetic data generation. Adversarial learning, as seen in SRecGAN~\cite{SRecGAN} and ECGAN-Rec~\cite{ECGAN-Rec}, introduces adversarial training to improve recommendation diversity and robustness, but may struggle with instability due to the discrete nature of item sequences. Overall, different methods balance trade-offs between temporal coherence, data efficiency, and computational complexity inherent to sequential recommendation.

\subsection{Social Recommendation}

\begin{table*}[!tb]
    \centering
    \small
    \caption{A summary of SSL recommendation methods in \textbf{Social Recommendation} and \textbf{Knowledge-aware Recommendation}.}
    \vspace{-0.15in}
    \begin{tabular}{c|c|c|c|c|c|c}    
        \toprule
        \multicolumn{7}{c}{\textbf{Social Recommendation (SocRec)}} \\
        \midrule
        Category & \multicolumn{3}{c|}{Method Information} & \multicolumn{3}{c}{Self-supervised Learning Paradigm} \\
        \midrule
        \multirow{10}{*}{CL} & Method & Year & Venue & View Creation & Pair Sampling & Contrastive Objective \\
        \cmidrule{2-7}
        & KCGN~\cite{KCGN}            & 2021  & AAAI          & Model-based           & Natural           & JS-based\\
        & SMIN~\cite{SMIN}            & 2021  & CIKM          & Model-based           & Natural           & JS-based\\
        & SEPT~\cite{SEPT}            & 2021  & KDD           & Data-based            & Score-based       & InfoNCE-based\\
        & MHCN~\cite{MHCN}            & 2021  & WWW           & Data-based            & Natural           & Explicit\\
        & DcRec~\cite{DcRecsocial}    & 2022  & CIKM          & Data-based            & Natural           & InfoNCE-based\\
        & SDCRec~\cite{SDCRec}        & 2022  & SIGIR         & Model-based           & Score-based       & InfoNCE-based\\
        & DUAL~\cite{DUAL}            & 2022  & TCSS          & Data-based            & Score-based       & InfoNCE-based\\
        & DSL~\cite{DSL}              & 2023  & IJCAI         & Model-based           & Natural           & Explicit\\
        & ReACL~\cite{ReACL}          & 2023  & Inf. Sci.     & Data-based            & Score-based       & InfoNCE-based\\
        & HGCL~\cite{HGCL}            & 2023  & WSDM          & Model-based           & Natural           & InfoNCE-based\\
        \midrule
        \multirow{5}{*}{AL} & Method & Venue & Year & \multicolumn{2}{c|}{Adversarial Learning Paradigm} & Discrimination Target \\
        \cmidrule{2-7}
        & Adit \textit{et al.}~\cite{krishnan2019modular}  & 2019    & CIKM          & \multicolumn{2}{c|}{Non-differentiable} & Sampled User-User Pair \\
        & RSGAN~\cite{RSGAN}                      & 2019    & ICDM          & \multicolumn{2}{c|}{Differentiable}     & Generated Item \\
        & DASO~\cite{DASO}                        & 2019    & IJCAI         & \multicolumn{2}{c|}{Non-differentiable} & Sampled User-User/Item Pair\\
        & ESRF~\cite{ESRF}                        & 2022    & TKDE          & \multicolumn{2}{c|}{Differentiable} & Sampled Neighbor Users \\
        \midrule
        \multicolumn{7}{c}{\textbf{Knowledge-aware Recommendation (KnoRec)}} \\
        \midrule
        Category & \multicolumn{3}{c|}{Method Information} & \multicolumn{3}{c}{Self-supervised Learning Paradigm} \\
        \midrule
        \multirow{9}{*}{CL} & Method & Year & Venue & View Creation & Pair Sampling & Contrastive Objective \\
        \cmidrule{2-7}
        & CKER~\cite{CKER}         & 2022    & Mathematics    & Model-based & Natural     & InfoNCE-based \\
        & KGIC~\cite{KGIC}         & 2022    & CIKM           & Data-based  & Score-based & InfoNCE-based \\
        & KGCL~\cite{KGCL}         & 2022    & SIGIR          & Data-based  & Natural     & InfoNCE-based \\
        & MCCLK~\cite{MCCLK}       & 2022    & SIGIR          & Model-based & Natural     & InfoNCE-based \\
        & KRec-C2~\cite{KRec-C2}   & 2023    & DASFAA         & Model-based & Natural     & InfoNCE-based \\
        & ML-KGCL~\cite{ML-KGCL}   & 2023    & DASFAA         & Model-based & Natural     & InfoNCE-based \\
        & HiCON~\cite{HiCON}       & 2023    & ICME           & Data-based  & Natural     & InfoNCE-based \\
        & KACL~\cite{KACL}         & 2023    & WSDM           & Model-based & Natural     & InfoNCE-based \\
        & HEK-CL~\cite{HEK-CL}     & 2025    & TOIS           & Data \& Feature-based & Natural & Explicit \\
        \midrule
        \multirow{3}{*}{RL} & Method & Venue & Year & \multicolumn{2}{c|}{Reconstructive Learning Paradigm} & Reconstruction Target \\
        \cmidrule{2-7}
        & KGRec~\cite{KGRec}       & 2023    & KDD            & \multicolumn{2}{c|}{Maksed Autoencoding} &  Triplets in the KG\\
        & DiffKG~\cite{DiffKG}     & 2024    & WSDM           & \multicolumn{2}{c|}{Denoising Diffusion} &  Triplets in the KG\\
    \bottomrule
    \end{tabular}\label{tab:social & knowledge}
    \vspace{-0.1in}
\end{table*}

\subsubsection{\textbf{Task Formulation}}
In social recommendation (SocRec), recommender systems are enhanced with side information that unveils the social relationships among users. The auxiliary observed data $\mathcal{X}$ in social recommendation is commonly represented as a user-user interaction graph $\mathcal{G}_{soc}$. In this graph, each interaction indicates the existence of a social relationship (\eg, friendship) between two users. The objective in SocRec is similar to general collaborative filtering: to recommend unexplored items to users by considering both collaborative and social information.

\subsubsection{\textbf{Contrastive Learning in SocRec}} Contrastive learning has become a significant approach in social recommendation, as demonstrated in Table~\ref{tab:social & knowledge}. By leveraging social relationship data, various contrastive views are generated, and mutual information is optimized through carefully designed pair-sampling strategies and objective functions.
\begin{itemize}
    \item \textbf{View Creation} in SocRec with CL can be categorized as data-based or model-based. The model-based approach emphasizes designing neural modules to encode features from social and collaborative data for contrastive views. Examples include KCGN~\cite{KCGN} using behavior and temporal-aware networks, and SMIN~\cite{SMIN} and SDCRec~\cite{SDCRec} employing heterogeneous GNNs while considering social relationships. DSL~\cite{DSL} utilizes GNNs for encoding distinct social-aware and collaborative-aware user embeddings. Conversely, data-based methods augment original graphs to generate diverse context-specific data. SEPT~\cite{SEPT} creates three graph views (preference, friend, and sharing) from user-item and user-user graphs, while MHCN~\cite{MHCN} generates hypergraphs among users using triangle motifs and applies hypergraph convolution. DcRec~\cite{DcRecsocial} diversifies views through edge addition and node/edge dropout, and ReACL~\cite{ReACL} constructs additional graphs using relationship-aware augmentation. In HGCL~\cite{HGCL}, a personalized knowledge transfer meta network encodes enhanced user and item representations for one view, while a lightweight GCN encodes another view for users/items.
    
    \item \textbf{Pair Sampling} in SocRec is designed based on the encoded nature of the created views. In natural sampling methods, a user or item may have multiple views. For instance, KCGN and MHCN pair node embeddings with their encoded sub-graph or hypergraph embeddings, considering the local-global relationship. Furthermore, several methods~\cite{SMIN, DcRecsocial, DSL, HGCL} encode multiple views for each node using graph augmentation or neural model design, treating the views of the same node as positive pairs. For score-based sampling, SEPT, SDCRec and ReACL label positive pairs by considering the representation similarly, and DUAL pre-computed a link-score for each edge in the graph based on the degree of the node, which is then utilized to choose positive pairs.
    
    \item \textbf{Contrastive Objective} in these methods commonly relies on InfoNCE-based~\cite{SEPT, DcRecsocial, SDCRec, DUAL, ReACL, HGCL} or JS-based objectives~\cite{KCGN, SMIN}. Notably, ReACL~\cite{ReACL} introduces an Aug-InfoNCE contrastive loss, leveraging an expanded set of positive samples to improve consistency among similar nodes and enhance the framework's generalization capability. In explicit optimization methods, MHCN employs a pair-wise ranking loss to maximize the mutual information between representations within a pair. On the other hand, DSL utilizes hinge loss to optimize the pair score, which proves beneficial in reducing the negative influence caused by noisy edges.
\end{itemize}

\subsubsection{\textbf{Adversarial Learning in SocRec}} Adversarial learning in social recommendation, as demonstrated in Table~\ref{tab:social & knowledge}, is frequently utilized to extract clean and representative social relations from the user-user graph in SocRec.
\begin{itemize}
    \item \textbf{Adversarial Learning Paradigm} in SocRec systems has emerged as a promising approach to leverage both user preferences and social relationships through generative adversarial frameworks. Adit \textit{et al.}~\cite{krishnan2019modular} proposed a modular adversarial approach to prevent model collapse in recommender systems with user-level social links, utilizing generative adversarial learning and a discriminator to differentiate sampled user pairs from interest and social domains. The parameter optimization involved policy gradient due to non-differentiable sampling operations. DASO~\cite{DASO} similarly leveraged sampled user-user or user-item pairs from the generator with policy gradient optimization. RSGAN~\cite{RSGAN} uses Gumbel-softmax for differentiable sampling and generates items based on social links, where the discriminator assigns higher scores to ground truth and socially aware items following social BPR optimization. In ESRF~\cite{ESRF}, a concrete selector layer is further utilized to avoid non-differentiable sampling for neighbor users for adversarial discrimination.
    \item \textbf{Discrimination Target} in SocRec adversarial frameworks focuses on effectively integrating social and collaborative information through regularized discrimination mechanisms. Adit \textit{et al.}~\cite{krishnan2019modular}, RSGAN~\cite{RSGAN}, and ESRF~\cite{ESRF} adopt similar approaches by sampling user-user/item pairs from generators and utilizing discriminators to enhance sample quality. Notably,  RSGAN~\cite{RSGAN} specifically employs generated items for discrimination, improving the generator's capability to produce more accurate items based on observed ground truth data.
\end{itemize}

\subsubsection{\textbf{In-depth Analysis of SSL-based Social Recommendation}}In social recommendation, two primary techniques-contrastive learning, adversarial learning-exhibit distinct advantages and limitations. Contrastive learning, while effective in leveraging social relationships to enhance representation learning~\cite{KCGN, SMIN}, often requires complex view creation and pair sampling strategies that can introduce computational overhead and potential noise from social connections. Adversarial learning, on the other hand, excels in refining the quality of user interactions by employing generative models~\cite{RSGAN} to discern meaningful patterns in social data.

\subsection{Knowledge-aware Recommendation}

\subsubsection{\textbf{Task Formulation}}
Items in reality have diverse attributes and labels, which can form a knowledge graph (KG) when connected to corresponding items. Knowledge-aware recommendation (KnoRec) leverages this external knowledge to enhance recommendations, represented as $\mathcal{G}_{k} = (h, r, t)$, where $h$ and $t$ are knowledge entities, and $r$ is their semantic relationship (e.g., \textit{book}, \textit{is written by}, \textit{author}). The knowledge graph $\mathcal{G}_{k}$ serves as external information to improve collaborative filtering performance, similar to social recommendation.

\subsubsection{\textbf{Contrastive Learning in KnoRec}} Contrastive learning in knowledge-aware recommendation, as shown in Table~\ref{tab:social & knowledge} merges knowledge graphs with user-item graphs, employing various encoding, graph construction, and augmentation techniques to generate diverse perspectives for improved performance.
\begin{itemize}
    \item \textbf{View Creation} in KnoRec involves the exploration of leveraging KGs in conjunction with the existing user-item graph to construct diverse perspectives. CKER~\cite{CKER} employs light-weight graph convolution and relation-aware convolution to encode interaction-aware and knowledge-aware item representations as two distinct views. KGIC~\cite{KGIC} utilizes the user-item interaction graph and KG to construct local and non-local graphs, encoding different views for nodes based on these two graphs. KGCL~\cite{KGCL} and ML-KGCL~\cite{ML-KGCL} both employ graph augmentation, such as edge dropout, twice to create contrastive views. MCCLK~\cite{MCCLK} first builds an item-item graph as the semantic view using relation-aware GNN, and then encodes collaborative and semantic representations for each node for contrastive learning. KRec-C2~\cite{KRec-C2} incorporates a category-level aggregation representation layer to obtain category-level signal features, which are contrasted with the original category features derived from the embedding function. HiCON~\cite{HiCON} leverages meta-paths to construct a high-order graph that encodes representations of users and items, contrasting them with representations obtained from the original interaction graph. KACL~\cite{KACL} encodes the interaction view using an augmented user-item graph and KG-aware item representations based on the KG, providing another view for contrastive learning.
    \item \textbf{Pair Sampling} in KnoRec contrastive learning methods focuses on constructing meaningful positive and negative sample pairs from knowledge graph-enhanced recommendation data. Generally, these methods create multiple views for each user and item node using various techniques~\cite{KGIC, KGCL, KACL}, which naturally form positive pairs for contrastive learning. KGIC~\cite{KGIC} uses node distances on local/non-local graphs to select positive pairs, enabling inter-graph and intra-graph interactive contrastive learning.
    \item \textbf{Contrastive Objective} in KnoRec systems aims to maximize the mutual information between different knowledge-enhanced views through InfoNCE-based optimization. All mentioned methods optimize the mutual information between different views using the InfoNCE-based lower bound as their contrastive objective.
\end{itemize}

\subsubsection{\textbf{Reconstructive Learning in KnoRec}} Reconstructive learning also emerges as a novel paradigm to enhance the performance of KnoRec (Table~\ref{tab:social & knowledge}). The intrinsic idea behind generative learning is to discover valuable knowledge triplets through generative tasks while mitigating the negative impact~\cite{K-BERT, TransA} caused by noisy or irrelevant knowledge.
\begin{itemize}
    \item \textbf{Reconstructive Learning Paradigm} in KnoRec encompasses both masked autoencoding and denoising diffusion approaches. In KGRec~\cite{KGRec}, a criterion score is initially computed for each knowledge triplet to identify the most valuable rationale triplets. Subsequently, the masking-and-reconstructing paradigm is employed to reconstruct these important triplets using node embeddings in the KG. This enables the encoding of information brought by these crucial triplets into the node representation, thereby enhancing recommendation performance. DiffKG~\cite{DiffKG} employs the concept of denoising diffusion by adding Gaussian noise to the triplets in the KG and subsequently removing it. This approach effectively eliminates the inherent noise present in the knowledge graph, providing a cleaner and augmented KG that can be utilized for inference and recommendation purposes.
    \item \textbf{Reconstruction Target} in both KGRec and DiffKG encompasses all the triplets present in the knowledge graph. This is because the original interactions within the KG may contain significant amounts of noise and redundant information that are irrelevant for recommendation~\cite{KGCL}. The generative learning process effectively generates valuable interactions within the KG through self-supervised learning, thereby achieving denoising.
\end{itemize}

\subsubsection{\textbf{In-depth Analysis of SSL-based Knowledge-aware Recommendation}} Knowledge-aware recommendation leverages SSL to address the challenges of sparse interactions and noisy knowledge integration. Contrastive learning methods, such as KGCL~\cite{KGCL} and KGIC~\cite{KGIC}, exploit multi-view representations derived from knowledge and interaction graphs, enhancing item/user embeddings by maximizing mutual information between views. Reconstructive approaches, like KGRec~\cite{KGRec} and DiffKG~\cite{DiffKG}, focus on reconstructing or denoising knowledge triplets, implicitly filtering noise and reinforcing valuable structural patterns. Their performance relies on the quality of selected reconstruction targets, and care must be taken to avoid underperformance when critical triplets are sparse or ambiguous.

\subsection{Cross-domain Recommendation}

\begin{table*}[!tb]
    \centering
    \small
    \caption{A summary of SSL recommendation methods in \textbf{Cross-domain Recommendation}.}
    \vspace{-0.15in}
    \begin{tabular}{c|c|c|c|c|c|c}    
        \toprule
        \multicolumn{7}{c}{\textbf{Cross-domain Recommendation (CroRec)}} \\
        \midrule
        Category & \multicolumn{3}{c|}{Method Information} & \multicolumn{3}{c}{Self-supervised Learning Paradigm} \\
        \midrule
        \multirow{11}{*}{CL} & Method & Year & Venue & View Creation & Pair Sampling & Contrastive Objective \\
        \cmidrule{2-7}
        & PCRec~\cite{PCRec}       & 2021 & CogMI & Data-based & Natural & InfoNCE-based \\
        & C2DSR~\cite{C2DSR}     & 2022 & CIKM      & Data-based            & Natural & JS-based \\
        & SASS~\cite{SASS}       & 2022 & CIKM      & Model-based           & Natural & InfoNCE-based \\
        & CDRIB~\cite{CDRIB}     & 2022 & ICDE      & Model-based           & Natural & JS-based \\
        & CLCDR~\cite{CLCDR}     & 2022 & ICONIP    & Data-based            & Natural & InfoNCE-based \\
        & CCDR~\cite{CCDR}       & 2022 & KDD       & Data \& Model-based   & Natural & InfoNCE-based \\
        & SITN~\cite{SITN}       & 2023 & AAAI      & Model-based           & Natural & InfoNCE-based \\
        & CATCL~\cite{CATCL}     & 2023 & DASFAA    & Feature-based         & Natural & InfoNCE-based \\
        & DCCDR~\cite{DCCDR}     & 2023 & DASFAA    & Model-based           & Natural & InfoNCE-based \\
        & DR-MTCDR~\cite{DR-MTCDR}  & 2023 & TOIS      & Data-based            & Natural & InfoNCE-based \\
        \midrule
        \multirow{4}{*}{RL} & Method & Venue & Year & \multicolumn{2}{c|}{Reconstructive Learning Paradigm} & Reconstruction Target \\
        \cmidrule{2-7}
        & RA/SA-VAE~\cite{RASA-VAE}  & 2021  & RecSys         & \multicolumn{2}{c|}{Variational Autoencoding}      & User Rating to Items \\
        & VDEA~\cite{VDEA}           & 2022  & SIGIR          & \multicolumn{2}{c|}{Variational Autoencoding}      & User Rating to Items \\
        & DiffCDR~\cite{DiffCDR}     & 2024  & arXiv          & \multicolumn{2}{c|}{Denoising Diffusion}           & User Feature \\
        \midrule
        \multirow{6}{*}{AL} & Method & Venue & Year & \multicolumn{2}{c|}{Adversarial Learning Paradigm} & Discrimination Target \\
        \cmidrule{2-7}
        & Su \textit{et al.}~\cite{su2022cross}         & 2022    & CIKM          & \multicolumn{2}{c|}{Differentiable}     & Domain Item Feature \\
        & RecGURU~\cite{RecGURU}               & 2022    & WSDM          & \multicolumn{2}{c|}{Differentiable}     & Domain User Feature \\
        & ACLCDR~\cite{ACLCDR}                 & 2023    & TKDD          & \multicolumn{2}{c|}{Non-Differentiable} & Augmented Interaction Matrix\\
        & DA-CDR~\cite{DA-CDR}                 & 2023    & TKDE          & \multicolumn{2}{c|}{Differentiable}     & Domain User/Item Feature\\
        & DA-DAN~\cite{DA-DAN}                 & 2023    & TOIS          & \multicolumn{2}{c|}{Differentiable}     & Domain User Feature\\
    \bottomrule
    \end{tabular}\label{tab:crossdomain}
    \vspace{-0.1in}
\end{table*}

\subsubsection{\textbf{Task Formulation}} Cross-domain recommendation (CroRec) transfers learned user preferences from a source domain to improve recommendations in a target domain, each containing different items based on domain-specific criteria. The items are divided into source domain set $\mathcal{V}_{s}$ and target domain set $\mathcal{V}_{t}$, with interaction matrices $\mathcal{A}_{s} \in \mathbb{R}^{|\mathcal{U}| \times |\mathcal{V}_{s}|}$ and $\mathcal{A}_{t} \in \mathbb{R}^{|\mathcal{U}| \times |\mathcal{V}{t}|}$, where $|\mathcal{U}|$ represents user count. Recommendations for users with uninteracted items ($v \in \mathcal{V}_{t}$) are made using a function $f$ that calculates preference scores ($y_{u, v}$) based on collaborative information across domains. CroRec can also involve multiple domains and be enhanced by incorporating sequential information~\cite{C2DSR, SITN}.

\subsubsection{\textbf{Contrastive Learning in CroRec}} Contrastive learning is crucial in SSL for cross-domain recommendation methods (Table~\ref{tab:crossdomain}), as it enables knowledge transfer from the target to the source domain by treating user interaction outcomes from different domains as distinct views. This section investigates diverse contrastive view creation methods using data from different domains and techniques for pair sampling and optimization in contrastive learning.
\begin{itemize}
    \item \textbf{View Creation} in CroRec contrastive learning methods encompasses diverse strategies to construct meaningful representations across different domains through data-based, model-based, and feature-based approaches. For data-based methods, PCRec~\cite{PCRec} uses random walks on source domain graphs, C2DSR~\cite{C2DSR} creates single-domain and cross-domain sequences with pooled prototype embeddings, CLCDR~\cite{CLCDR} replaces BPR loss with contrastive loss based on interaction data, DR-MTCDR~\cite{DR-MTCDR} employs node/edge dropping, and CCDR~\cite{CCDR} uses sub-graph augmentation for intra-domain CL. For model-based approaches, SASS~\cite{SASS} and CDRIB~\cite{CDRIB} maximize differences between views of identical nodes, CCDR~\cite{CCDR} uses neural-encoded representations from different domains for inter-domain CL, SITN~\cite{SITN} encodes user sequences from both domains with clustering techniques, and DCCDR~\cite{DCCDR} projects representations into domain-specific and domain-invariant forms. For feature-based methods, CATCL~\cite{CATCL} adopts a feature-based view creation approach, similar to SimGCL, by adding noise to the node features during inference, which generates diverse views that can be utilized for learning purposes.
    \item \textbf{Pair Sampling} relies on natural sampling in these works, where multiple views are created for each data object. Positive pairs are formed by two views of the same object, while views from other objects serve as negative samples. Encoded representations from multiple views for the same user/item node can be paired, originating from different domains~\cite{SASS, CDRIB, CCDR} or generated through data augmentation techniques~\cite{PCRec, CATCL}. Additionally, these representations can be encoded by specifically designed neural modules~\cite{PCRec, DCCDR}.
    \item \textbf{Contrastive Objective} within CL-based cross-domain recommendation methods typically employs the InfoNCE-based objective to optimize the mutual information between positive views~\cite{PCRec, SASS, CLCDR, CCDR, SITN, CATCL, DCCDR, DR-MTCDR}. Additionally, some methods utilize the JS-based lower bound for optimization. For instance, C2DSR~\cite{C2DSR} corrupts the prototype representation to generate negative views, while CDRIB~\cite{CDRIB} derives the JS-based contrastive term based on the information bottleneck regularization theory.
\end{itemize}

\subsubsection{\textbf{Reconstructive Learning in CroRec}} In CroRec, generative learning (as shown in Table~\ref{tab:crossdomain}) employs variational autoencoding and diffusion models to align and transfer knowledge between the source and target domains. The core idea is to facilitate knowledge sharing and adaptation across domains.

\begin{itemize}
    \item \textbf{Reconstructive Learning Paradigm} in CroRec begins by employing deep generative latent variable models (\ie, variational auto-encoders) to encode the latent space of user preferences. Subsequently, it transitions to the denoising diffusion model to facilitate knowledge transfer across domains. Salah \textit{et al.} propose RA/SA-VAE~\cite{RASA-VAE} to simultaneously fit the target observations and align the hidden space encoded by VAE with the source latent space. They introduce both rigid alignment and soft alignment techniques with varying degrees of user preference alignment. Furthermore, VDEA~\cite{VDEA} extends the alignment of latent spaces using variational autoencoding at both local and global levels, allowing for the exploitation of domain-invariant features across different domains, including both overlapped and non-overlapped users. Recently, DiffCDR~\cite{DiffCDR} leveraged denoising probabilistic models (DPMs) to process noisy data and generate denoised results, effectively transferring data between domains.
    \item \textbf{Reconstruction Target} in cross-domain recommendation systems serves as a fundamental mechanism to bridge information gaps between domains by reconstructing missing or incomplete data representations. RA/SA-VAE and VDEA both adopt the user rating reconstruction paradigm to optimize the variational model. In these models, the reconstruction target is the user ratings for items, represented as vectors in the model. The obtained ratings include ratings for uninteracted items, which are subsequently used for ranking and recommendation. On the other hand, DiffCDR focuses on reconstructing user features in the target domain by reversing the diffusion process conditioned on the corresponding user's information in the source domain.
\end{itemize}

\subsubsection{\textbf{Adversarial Learning in CroRec}} Adversarial learning in Table~\ref{tab:crossdomain} employs adversarial domain adaptation to generate domain-independent features, encoding domain-invariant user interaction preferences for effective knowledge transfer. Some methods also utilize adversarial samples to improve model learning.

\begin{itemize}
    \item \textbf{Adversarial Learning Paradigm} in CroRec leverages discriminative mechanisms to facilitate effective domain knowledge transfer. Several methods employ adversarial learning for cross-domain recommendation. For instance, Su \textit{et al.}~\cite{su2022cross} use a discriminator to identify domain-specific features, with gradients propagating to optimize the feature generator. Similarly, RecGURU~\cite{RecGURU} feeds encoded user representations from both domains into a discriminator for classification, enabling differentiable adversarial learning. Moreover, ACLCDR~\cite{ACLCDR} utilizes Deep Q-Learning to design a generator that enhances the interaction matrix with fake interactions, optimizing it using reward signals from downstream tasks. In addition, DA-CDR~\cite{DA-CDR} encodes domain features of users and items to fool the discriminator, ensuring effective domain knowledge transfer through adversarial learning using a gradient reversal layer (GRL) for differentiable training. Furthermore, DA-DAN~\cite{DA-DAN} incorporates adversarial domain adaptation into unsupervised non-overlapping CroRec, with the discriminator classifying the domain of encoded user features for overall differentiable training.
    \item \textbf{Discrimination Target} in CroRec focuses on distinguishing domain-specific features to promote domain-invariant representation learning.. Su \textit{et al.}~\cite{su2022cross}'s domain adaptation method uses features from source and target domains to discriminate between them, forcing the generator to encode domain-invariant user preferences, thereby confusing the discriminator. RecGURU adopts a similar approach, discriminating encoded user representations from both domains to generalize user representation learning across domains. In ACLCDR, the augmented interaction matrix from DQN is used for inference in subsequent models, with the recommendation model acting as an implicit discriminator. DA-CDR employs a discriminator to classify domain features of users/items, encouraging the generator to encode additional domain-shared information for cross-domain recommendation. Lastly, DA-DAN encodes user-specific item sequence representations as domain features for the discriminator's classification task, enabling the learning of domain-invariant features for recommendation.
\end{itemize}

\subsubsection{\textbf{In-depth Analysis of SSL-based Cross-domain Recommendation}}In cross-domain recommendation, three main SSL approaches-contrastive learning, generative learning, and adversarial learning-each present unique strengths. Contrastive learning is effective in generating multiple views~\cite{PCRec, SASS} and enhancing mutual information across different domains, facilitating smooth knowledge transfer. Reconstructive learning, utilizing frameworks like variational autoencoders, aids in aligning latent spaces and reconstructing user preferences, offering a comprehensive representation of interactions~\cite{RASA-VAE, VDEA}. Yet, it may face challenges in scalability with high-dimensional datasets. Adversarial learning focuses on producing domain-invariant features~\cite{su2022cross, RecGURU} that improve generalization across domains, but its training can be unstable due to the competitive nature of the adversarial process involved.

\subsection{Bundle Recommendation}

\begin{table*}[!tb]
    \centering
    \small
    \caption{A summary of SSL recommendation methods in \textbf{Bundle Recommendation} and \textbf{Group Recommendation}.}
    \vspace{-0.15in}
    \begin{tabular}{c|c|c|c|c|c|c}    
        \toprule
        \multicolumn{7}{c}{\textbf{Bundle Recommendation (BunRec)}} \\
        \midrule
        Category & \multicolumn{3}{c|}{Method Information} & \multicolumn{3}{c}{Self-supervised Learning Paradigm} \\
        \midrule
        \multirow{3}{*}{CL} & Method & Year & Venue & View Creation & Pair Sampling & Contrastive Objective \\
        \cmidrule{2-7}
        & MIDGN~\cite{MIDGN}            & 2022  & AAAI         & Model-based             & Natural & InfoNCE-based \\
        & CrossCBR~\cite{CrossCBR}      & 2022  & KDD          & Data \& Feature-based   & Natural & InfoNCE-based \\
        \midrule
        \multirow{2}{*}{RL} & Method & Venue & Year & \multicolumn{2}{c|}{Reconstructive Learning Paradigm} & Reconstruction Target \\
        \cmidrule{2-7}
        & DGMAE~\cite{DGMAE}       & 2023    & SIGIR            & \multicolumn{2}{c|}{Maksed Autoencoding} &  Edge in User-Item Graph\\
        \midrule
        \multicolumn{7}{c}{\textbf{Group Recommendation (GroRec)}} \\
        \midrule
        Category & \multicolumn{3}{c|}{Method Information} & \multicolumn{3}{c}{Self-supervised Learning Paradigm} \\
        \midrule
        \multirow{5}{*}{CL} & Method & Year & Venue & View Creation & Pair Sampling & Contrastive Objective \\
        \cmidrule{2-7}
        & GroupIM~\cite{GroupIM}    & 2020  & SIGIR         & Model-based           & Natural           & JS-based \\
        & HHGR~\cite{HHGR}          & 2021  & CIKM          & Data-based            & Natural           & JS-based \\
        & CubeRec~\cite{CubeRec}    & 2022  & SIGIR         & Model-based           & Score-based       & Explicit\\
        & SGGCF~\cite{SGGCF}        & 2023  & WSDM          & Model-based           & Natural           & InfoNCE-based\\
    \bottomrule
    \end{tabular}\label{tab:bundle & group}
    \vspace{-0.1in}
\end{table*}

\subsubsection{\textbf{Task Formulation}} In Bundle Recommendation (BunRec), the auxiliary observed data $\mathcal{X}$ includes the item-bundle affiliation, where a bundle $b \in \mathcal{B}$ represents a set of items for recommendation. Therefore, we have two interaction matrices: the user-bundle interaction matrix $\mathcal{A}_{U-B} \in \mathbb{R}^{|\mathcal{U}| \times |\mathcal{B}|}$ and the user-item interaction matrix $\mathcal{A}_{U-I} \in \mathbb{R}^{|\mathcal{U}| \times |\mathcal{I}|}$. Additionally, we use the matrix $\mathcal{A}_{I-B} \in \mathbb{R}^{|\mathcal{V}| \times |\mathcal{B}|}$ to record the item-bundle affiliation. The overall objective is to recommend uninteracted bundles to each user by predicting their preference scores $y_{u,b}$.

\subsubsection{\textbf{Contrastive Learning in BunRec}} In bundle recommendation with CL (Table~\ref{tab:bundle & group}), the intricate relationships between items and bundles, along with user-item interactions, are leveraged to construct user-bundle relationships. Various methodologies are subsequently designed to create contrastive views based on these multi-level relationships.
\begin{itemize}
    \item \textbf{View Creation} in BunRec contrastive learning methods focuses on constructing meaningful representations from multi-granular bundle-item relationships and user interaction patterns. MIDGN~\cite{MIDGN} employs neural models to encode distinct representations for each user and bundle, treating them as separate views, while incorporating a graph disentangle module to encode intent-aware representations and utilizing a user-bundle graph with LightGCN for cross-view representation learning. CrossCBR~\cite{CrossCBR} constructs the user-bundle graph by leveraging user-item and bundle-item relationships, subsequently encoding user and bundle representations through graph augmentation and embedding augmentation techniques in both bundle and item views. Consequently, each user and bundle node possesses dual views for contrastive learning, following a similar paradigm to MIDGN.
    
    \item \textbf{Pair Sampling} in BunRec contrastive methods leverages the dual-view architecture to construct positive and negative sample pairs for effective contrast learning. Both MIDGN and CrossCBR encode two distinct views for each user and bundle in the dataset, where the dual views of a specific user/bundle node naturally constitute the positive pair, while views from different nodes serve as negative samples for contrastive comparison.
    
    \item \textbf{Contrastive Objective} in BunRec systems aims to optimize multi-granular representations through self-supervised learning signals derived from bundle-item hierarchies. Both MIDGN and CrossCBR utilize the InfoNCE-based objective to optimize their models through self-supervised learning mechanisms.
\end{itemize}

\subsubsection{\textbf{Reconstructive Learning in BunRec}} In BunRec, generative learning (Table~\ref{tab:bundle & group}) adopts auto-encoder paradigm to learn meaningful representations. The primary approach (\ie, DGMAE~\cite{DGMAE}) is illustrated as follows.
\begin{itemize}
    \item \textbf{Reconstructive Learning Paradigm} leverages knowledge distillation combined with masked autoencoding mechanisms. In DGMAE, a teacher GNN model is first trained using user-bundle interactions, with its knowledge subsequently distilled into a student graph-masked auto-encoder model following mask autoencoding principles.
    
    \item \textbf{Reconstruction Target} here focuses on masking and reconstructing critical structures to enhance model robustness. DGMAE employs an masking strategy on edges within the graph, prioritizing edges with lower sparsity scores for masking to increase reconstruction challenge and improve model robustness.
\end{itemize}

\subsubsection{\textbf{In-depth Analysis of SSL-based Bundle Recommendation}} SSL techniques in this scenario address data sparsity and complex item-bundle synergies through distinct paradigms. CL-based methods~\cite{MIDGN, CrossCBR} enhance representation alignment across user-bundle and item-bundle views, improving robustness to data sparsity through multi-perspective augmentation. However, their effectiveness depends on domain-informed view design to balance semantic relevance and computational efficiency. Reconstructive approaches~\cite{DGMAE} systematically recover masked user-item interactions, prioritizing high-affinity patterns to alleviate sparsity. A integration of them in the future could holistically address unique demands, such as capturing transient user intents and evolving bundle-item synergies.

\subsection{Group Recommendation}

\subsubsection{\textbf{Task Formulation}} The goal of Group Recommendation (GroRec) is to recommend items to a group denoted as $o \in \mathcal{O}$, which represents a set of users. In addition to user-item interactions, there are also group-item interactions represented by $\mathcal{A}_{O-I} \in \mathbb{R}^{|\mathcal{O}| \times |\mathcal{I}|}$ and group-user affiliations represented by $\mathcal{A}_{O-U} \in \mathbb{R}^{|\mathcal{O}| \times |\mathcal{U}|}$. The recommender model will predict a score $y_{o,v}$ for each group-item pair $(o, v)$ for recommendation.

\subsubsection{\textbf{Contrastive Learning in GroRec}} In GroRec, contrastive learning is vital in SSL methods (Table~\ref{tab:bundle & group}), utilizing diverse user-item and user-group relationships to create multiple views for each node, enabling effective training.
\begin{itemize}
    \item \textbf{View Creation} in GroRec leverages multi-granular group structures and user-item interactions to construct meaningful representation views. GroupIM~\cite{GroupIM} pioneers the use of mutual information maximization, encoding user preference and group representations as views for contrastive learning. Meanwhile, HHGR~\cite{HHGR} constructs fine- and coarse-grained user-level hypergraphs, encoding augmented views of user nodes at different granularities. Moreover, CubeRec~\cite{CubeRec} transforms the recommendation scenario using a hypercube framework, encoding user and item representations with LightGCN and treating each user representation as a view for pairing and learning. Lastly, SGGCF~\cite{SGGCF} adopts a data-based view creation approach, constructing a group-user-item graph, augmenting it with node/edge dropout, and encoding diverse views for each node.
    \item \textbf{Pair Sampling} in GroRec constructs positive and negative sample pairs from group-user relationships and multi-view representations.. In BunRec, users are linked with user groups, forming positive pairs between user and group representations. GroupIM maximizes mutual information (MI) between these representations. In HHGR and SGGCF, each graph node has multiple augmented views, which are naturally treated as positive pairs, while views of other nodes are considered negative samples. CubeRec constructs a group hypercube based on user representations, identifying the intersection between two group hypercubes. Each overlapping user is paired with the hypercube as a positive pair, and other users are considered negative samples. The distance between each user and the hypercube is then calculated for contrastive optimization.
    \item \textbf{Contrastive Objective} employs various optimization strategies tailored to group-aware recommendation scenarios.. GroupIM utilizes a noise-contrastive objective with a binary cross-entropy loss, specifically a JS-based objective which maximizes the mutual information (MI) between positive pairs. On a related note, HHGR also employs a similar JS-based optimization approach. In CubeRec, an explicit margin loss is utilized to ensure that the distance-to-hypercube of positive pairs is smaller than that of negative pairs. For SGGCF, it adopts the InfoNCE-based objective to achieve the contrastive learning process. 
\end{itemize}

\subsubsection{\textbf{In-depth Analysis of SSL-based Group Recommendation}}Contrastive learning-based methods have emerged as the mainstream approach in group recommendation due to their ability to effectively leverage user-item and user-group relationships. These methods, such as GroupIM~\cite{GroupIM}, HHGR~\cite{HHGR}, and SGGCF~\cite{SGGCF}, excel in capturing fine-grained interaction patterns through multi-view representations, enabling robust modeling of the complex dynamics inherent in groups. However, CL-based methods are inherently constrained by their significant computational demands, which present a trade-off in fully optimizing group recommendation with CL. It would be promising to explore other SSL techniques, such as adversarial or reconstructive approaches, that may offer complementary advantages.

\subsection{Multi-behavior Recommendation}

\begin{table*}[!tb]
    \centering
    \small
    \caption{A summary of SSL recommendation methods in \textbf{Multi-behavior Recommendation} and \textbf{Multi-modal Recommendation}.}
    \vspace{-0.15in}
    \begin{tabular}{c|c|c|c|c|c|c}    
        \toprule
        \multicolumn{7}{c}{\textbf{Multi-behavior Recommendation (MbeRec)}} \\
        \midrule
        Category & \multicolumn{3}{c|}{Method Information} & \multicolumn{3}{c}{Self-supervised Learning Paradigm} \\
        \midrule
        \multirow{10}{*}{CL} & Method & Year & Venue & View Creation & Pair Sampling & Contrastive Objective \\
        \cmidrule{2-7}
        & HMG-CR~\cite{HMG-CR}      & 2021  & ICDM          & Data \& Model-based     & Natural                    & InfoNCE-based \\
        & S-MBRec~\cite{S-MBRec}    & 2021  & ICDM          & Model-based             & Score-based                & InfoNCE-based \\
        & MMCLR~\cite{MMCLR}        & 2022  & DASFAA        & Model-based             & Natural                    & Explicit \\
        & CML~\cite{CML}            & 2022  & WSDM          & Model-based             & Natural                    & InfoNCE-based \\
        & IICL~\cite{IICL}          & 2023  & DASFAA        & Model-based             & Natural \& Score-based     & InfoNCE-based \\
        & MixMBR~\cite{MixMBR}      & 2023  & DASFAA        & Model-based             & Natural                    & InfoNCE-based \\
        & TMCL~\cite{TMCL}          & 2023  & DASFAA        & Model-based             & Natural                    & InfoNCE-based \\
        & RCL~\cite{RCL}            & 2023  & RecSys        & Model-based             & Natural                    & InfoNCE-based \\
        & MBSSL~\cite{MBSSL}        & 2023  & SIGIR         & Data \& Model-based     & Natural                    & InfoNCE-based \\
        & KMCLR~\cite{KMCLR}        & 2023  & WSDM          & Data \& Model-based     & Natural                    & InfoNCE-based \\
        & MENTOR~\cite{MENTOR}      & 2025  & AAAI          & Data \& Feature-based   & Natural                    & InfoNCE-based \\
        \midrule
        \multirow{3}{*}{RL} & Method & Venue & Year & \multicolumn{2}{c|}{Reconstructive Learning Paradigm} & Reconstruction Target \\
        \cmidrule{2-7}
        & VCGAE~\cite{VCGAE}        & 2023    & ICDM            & \multicolumn{2}{c|}{Variational Autoencoding} &  Target Behavior Graph\\
        & BVAE~\cite{BVAE}          & 2023    & RecSys          & \multicolumn{2}{c|}{Variational Autoencoding} &  User Interaction Vector\\
        \midrule
        \multicolumn{7}{c}{\textbf{Multi-modal Recommendation (MmoRec)}} \\
        \midrule
        Category & \multicolumn{3}{c|}{Method Information} & \multicolumn{3}{c}{Self-supervised Learning Paradigm} \\
        \midrule
        \multirow{8}{*}{CL} & Method & Year & Venue & View Creation & Pair Sampling & Contrastive Objective \\
        \cmidrule{2-7}
        & Liu \textit{et al.}~\cite{liu2022multi}         & 2022  & ICMR        & Model-based           & Natural           & JS-based \\
        & CMI~\cite{CMI}            & 2022  & SIGIR       & Data-based            & Natural           & InfoNCE-based\\
        & MMGCL~\cite{MMGCL}        & 2022  & SIGIR       & Data-based            & Natural           & InfoNCE-based\\
        & SLMRec~\cite{SLMRec}      & 2022  & TMM         & Data-based            & Natural           & InfoNCE-based \\
        & MICRO~\cite{MICRO}        & 2023  & TKDE        & Model-based           & Natural           & InfoNCE-based \\
        & BM3~\cite{BM3}            & 2023  & WWW         & Feature-based         & Natural           & Explicit\\
        & MMSSL~\cite{MMSSL}        & 2023  & WWW         & Model-based           & Natural           & InfoNCE-based\\
        \midrule
        \multirow{2}{*}{RL} & Method & Venue & Year & \multicolumn{2}{c|}{Reconstructive Learning Paradigm} & Reconstruction Target \\
        \cmidrule{2-7}
        & MVGAE~\cite{MVGAE}        & 2022    & TMM            & \multicolumn{2}{c|}{Variational Autoencoding} &  Edge in Graph\\
        \midrule
        \multirow{2}{*}{AL} & Method & Venue & Year & \multicolumn{2}{c|}{Adversarial Learning Paradigm} & Discrimination Target \\
        \cmidrule{2-7}
        & MMSSL~\cite{MMSSL}        & 2023    & WWW            & \multicolumn{2}{c|}{Differentiable} &  User Interaction to Items\\
    \bottomrule
    \end{tabular}\label{tab:behavior & modal}
    \vspace{-0.1in}
\end{table*}

\subsubsection{\textbf{Task Formulation}} In Multi-behavior Recommendation (MbeRec), the user-item interaction is extended to incorporate behavior heterogeneity, resulting in a 3D tensor denoted as $\mathcal{A} \in \mathbb{R}^{|\mathcal{U}| \times |\mathcal{I}| \times |\mathcal{B}|}$, where $\mathcal{B} = \{b_1, ..., b_K \}$ depicts the set of different types of behaviors (\eg, page view, add-to-cart). The MbeRec aims to provide item recommendations for the target behavior (e.g., purchase) by leveraging the diverse behavior information.

\subsubsection{\textbf{Contrastive Learning in MbeRec}} In MbeRec, CL leverages multi-behavior user-item interactions to create behavior-specific views for effective representation learning. The methods are summarized in Table~\ref{tab:behavior & modal}.
\begin{itemize}
    \item \textbf{View Creation} in MbeRec focuses on constructing behavior-aware representations through model-based and data-level augmentation strategies. HMG-CR~\cite{HMG-CR} uses hyper meta-paths to construct multiple hyper meta-graphs with distinct encoders for behavior embeddings. S-MBRec~\cite{S-MBRec} partitions multi-behavior graphs into sub-graphs, encoding multiple GCN embeddings as views. MMCLR~\cite{MMCLR} encodes sequence and graph views under different behaviors, obtaining fusion views by combining representations. CML~\cite{CML} and IICL~\cite{IICL} leverage behavior-aware GNNs to encode user embeddings for each behavior type as separate views. MixMBR~\cite{MixMBR}, TMCL~\cite{TMCL}, and RCL~\cite{RCL} encode behavior-specific embeddings through parameter mixup, temporal-aware learning, and short/long-term interest modeling respectively. MBSSL~\cite{MBSSL} employs behavior-aware GNNs with edge dropout for augmented views, while KMCLR~\cite{KMCLR} incorporates knowledge-based data augmentation.
    
    \item \textbf{Pair Sampling} in MbeRec methods constructs positive and negative pairs from behavior-specific representations and similarity-based mechanisms. Most methods treat different behavior views of the same node as positive pairs, with other nodes serving as negative samples. HMG-CR pairs user behavior embeddings as positives, S-MBRec calculates interaction-based similarity scores for pairing, and MBSSL filters false negatives using user similarity scores. IICL employs both intra-behavior (even-layered embeddings) and inter-behavior (different behaviors for same user) contrastive learning, while RCL pairs fused multi-behavior views with behavior-specific views.
    
    \item \textbf{Contrastive Objective} in MbeRec systems primarily employs InfoNCE-based optimization to maximize behavior-aware representation learning. Most methods~\cite{HMG-CR, S-MBRec, CML, IICL, MixMBR, TMCL, RCL, KMCLR} utilize the InfoNCE objective for contrastive optimization. MMCLR explicitly pulls positive pairs closer while pushing negative pairs apart by ensuring higher similarity scores for positive pairs.
\end{itemize}

\subsubsection{\textbf{Reconstructive Learning in MbeRec}} Reconstructive learning in MbeRec (Table~\ref{tab:behavior & modal}) adopts the variational autoencoding paradigm to reconstruct multi-behavior interactions, thereby providing informative auxiliary learning signals for behavior-aware recommendation.
\begin{itemize}
    \item \textbf{Reconstructive Learning Paradigm} in MbeRec leverages variational autoencoding mechanisms to model behavior heterogeneity and capture complex interaction patterns. Both VCGAE~\cite{VCGAE} and BVAE~\cite{BVAE} adopt the variational auto-encoding generative paradigm for modeling behavior heterogeneity. VCGAE designs a variational graph autoencoder that utilizes auxiliary behavior fusion to obtain the variance vector and employs a behavior transfer network to derive the mean vector, jointly forming the distribution of latent factors. BVAE uses a behavior-aware semi-encoder to encode the variance vector based on users' historical interactions and learns the mean vector through a global feature filtering network, resulting in behavior-aware latent user representations.
    \item \textbf{Reconstruction Target} here focuses on reconstructing behavior-specific patterns to enhance representation learning. In the VAE paradigm, the obtained latent representations are fed into the decoder for generation tasks. VCGAE targets the reconstruction of edges in the graph for specific target behaviors (\eg, purchase), while BVAE focuses on generating comprehensive user interaction vectors across different behavioral modalities.
\end{itemize}

\subsubsection{\textbf{In-depth Analysis of SSL-based Multi-behavior Recommendation}} In multi-behavior recommendation, CL-based methods excel at modeling behavior heterogeneity by creating behavior-specific views and leveraging multi-behavior interactions~\cite{HMG-CR, MMCLR, MBSSL}. These methods effectively capture inter- and intra-behavior information, enabling fine-grained representation learning. Meanwhile, reconstructive methods~\cite{VCGAE, BVAE} offer an alternative by focusing on behavior reconstruction to uncover latent patterns. While CL-based methods dominate for their flexibility and multi-view learning, integrating generative paradigms could further improve the robustness of multi-behavior recommendation models in the future.

\subsection{Multi-modal Recommendation}

\subsubsection{\textbf{Task Formulation}} In the context of Multi-modal Recommendation (MmoRec), auxiliary observed data $\mathcal{X}$ contains item multi-modal information. Typically, an item can possess auxiliary features such as text descriptions, images, or acoustic data. Each item is associated with modality-specific features $\mathbf{e}_{m}$, where $m \in \mathcal{M}$ represents modalities. MmoRec aims to utilize this multi-modal information to improve recommendation accuracy.

\subsubsection{\textbf{Contrastive Learning in MmoRec}} Contrastive learning in MmoRec (Table~\ref{tab:behavior & modal}) cleverly leverages the multi-modal information of items to construct different contrastive views. This enables effective fusion of multi-modal information and promotes the learning of recommenders.
\begin{itemize}
    \item \textbf{View Creation} in MmoRec methods constructs meaningful representations from multi-modal item features through data-based, model-based, and feature-based strategies. In data-based view creation methods, CMI~\cite{CMI} conducts data-level item sequence augmentation and encodes user interest embeddings to generate contrastive views. MMGCL~\cite{MMGCL} employs modality masking and edge dropout for augmented graph views. Additionally, SLMRec~\cite{SLMRec} utilizes data-level augmentation techniques such as feature dropout and masking to obtain two views for each node. In contrast, model-based methods like Liu \textit{et al.}~\cite{liu2022multi} leverage a text/image encoder to encode two views for each item, which are then used for contrastive learning. Furthermore, MICRO~\cite{MICRO} mines the latent graph structure of items and utilizes graph neural networks to encode modality-specific views and generate a fused view for each item. Meanwhile, MMSSL~\cite{MMSSL} employs adversarial learning to obtain a modality-aware user-item graph and encodes modality-fused user embeddings as one view and another user's view based on collaborative information. For feature-based view creation, BM3~\cite{BM3} encodes modality embeddings, applies feature-level embedding dropout for augmented views, and includes an ID-based view from ID embeddings.
    \item \textbf{Pair Sampling} in MmoRec methods constructs positive and negative pairs from multi-modal representations and augmented views. All the methods follow natural sampling. In~\cite{liu2022multi}, positive pairs are formed by two views from different modalities of the same item, and negative pairs consist of views from different items. CMI creates positive pairs from two augmented interest-based views for the same user. MMGCL forms positive pairs from randomly augmented views and challenging negatives obtained by replacing modality data. SLMRec forms positive pairs with two data-augmented views of each node and ID-modality and other modality representations. MICRO maximizes agreement between item representations under individual modalities and fused multi-modal representations, forming positive pairs with modality-view and fused view of the same item and negative pairs with views from other items. BM3 aligns inter-modality by pairing augmented modality-views with the ID-based view and intra-modality by pairing modality-views of the same item. MMSSL forms positive pairs with the fused modality-view and collaborative-aware view of the same user, and negative pairs with views from other users.
    \item \textbf{Contrastive Objective} in MmoRec systems utilizes InfoNCE-based~\cite{CMI, MMGCL, SLMRec, MICRO, MMSSL} or JS-based~\cite{liu2022multi} loss functions, while BM3 directly aligns views with cosine similarity.
\end{itemize}

\subsubsection{\textbf{Reconstructive Learning in MmoRec}} Reconstructive learning, as recently shown by MVGAE~\cite{MVGAE} (Table~\ref{tab:behavior & modal}), employs the graph VAE to transform modality data into a latent space and utilizes generation tasks to guide the learning.
\begin{itemize}
    \item \textbf{Reconstructive Learning Paradigm} in MmoRec integrates multi-modal information through variational autoencoding, where MVGAE assigns mean and variance vectors to each user-item graph node and employs product-of-experts to fuse modality-specific latent representations.
    \item \textbf{Reconstruction Target} here reconstructs graph structural information, where MVGAE uses encoder-derived latent representations to reconstruct user-item interaction edges via inner product decoder and BPR loss.
\end{itemize}

\subsubsection{\textbf{Adversarial Learning in MmoRec}} In MmoRec, adversarial learning is recently used in MMSSL~\cite{MMSSL} (Table~\ref{tab:behavior & modal}) to perform structure learning based on multi-modal information, revealing modality-aware user-item interactions.
\begin{itemize}
    \item \textbf{Adversarial Learning Paradigm} in MmoRec generates user-item structures through multi-modal adversarial mechanisms. MMSSL generates a modality-aware user-item graph structure using a neural generator with item multi-modal features. Then, the user-specific interaction vector undergoes adversarial learning in a discriminator.
    \item \textbf{Discrimination Target} here distinguishes real and generated user-item interactions, where the user-item graph structure learned in MMSSL is split into user-specific interaction vectors, which are used for discrimination.
\end{itemize}

\subsubsection{\textbf{In-depth Analysis of SSL-based Multi-modal Recommendation}} Three SSL techniques each have distinct attributes. Contrastive learning methods~\cite{CMI, BM3} excel at utilizing diverse views of multi-modal data to enhance representation learning; however, they may face challenges in creating high-quality multi-modal views for optimization. Reconstructive methods~\cite{MVGAE} and adversarial learning approaches~\cite{MMSSL} both provide robust capabilities: the former captures complex interdependencies among modalities, while the latter offers a dynamic framework for modeling intricate user-item interactions, promoting more nuanced embeddings. Both approaches require the appropriate design of reconstruction objectives and adversarial processes to effectively harness multi-modal data.

%% file: future.tex
\section{Discussions and Future Directions}
\label{sec:future}
In this section, we aim to delve into several open problems and potential future directions in the field of self-supervised learning for recommendation systems. By exploring and analyzing these challenges and opportunities, we aim to stimulate and encourage further research, development, and innovation in this rapidly advancing field of study.

\subsection{Towards Foundation Recommender Models}
Foundation model~\cite{bommasani2021opportunities}, trained on massive datasets and exhibiting remarkable generalization capabilities to handle a wide range of downstream tasks, has garnered significant attention from researchers across various domains~\cite{radford2021learning, bubeck2023sparks, xia2024opengraph, GraphGPT}. In recommender systems, current models using SSL techniques have shown significant performance, but they are limited by evaluation settings focused on a single dataset. Recent research in CV and NLP has improved models' generalization abilities through self-supervised learning using contrastive learning~\cite{radford2021learning} or generative learning~\cite{touvron2023llama} on large amounts of data. A promising future direction is to design foundation recommender models that leverage SSL to learn user-item interaction patterns from massive data and achieve zero-shot cross-data reasoning and recommendation.

\subsection{Unleashing the Potential of Denoised Diffusion Models}
Denoised diffusion models have demonstrated remarkable proficiency in generating diverse types of data, including images, text, and even structured data. This new and popular generative learning paradigm has been effectively applied in various domains, as evidenced by recent studies such as~\cite{rombach2022high, gong2022diffuseq}, and~\cite{vignac2022digress}. Recent studies in recommendation with generative self-supervised learning have started leveraging diffusion models either to generate augmented data~\cite{DiffuASR} or as the backbone model for inference~\cite{DiffRec}. We believe that the impressive generation capabilities demonstrated by diffusion models will bring new insights and fascinating models to the generative learning-enhanced recommendation in the near future, making it a promising research line.

\subsection{Self-Supervised Integration of Large Language Models}
Large language models (LLMs) have gained significant attention due to their exceptional performance in various domains~\cite{zhao2023survey, HiGPT, LLaVA}. In recommender systems, LLMs can generate user/item profiles~\cite{ren2023representation, wei2023llmrec}, explain user-item interactions~\cite{li2023prompt}, and serve as the backbone for recommendation with instruction tuning~\cite{zhang2023recommendation}. The integration of large language models has brought about an unprecedented abundance of diverse, rich, and high-quality textual modality data. However, the challenge of effectively harnessing this textual capability remains an open research question. Recent works have employed self-supervised learning techniques, such as contrastive learning~\cite{ren2023representation} and mask-reconstruction~\cite{wei2023llmrec}, to align the knowledge of LLMs with recommenders. As we look to the future, self-supervised learning techniques are poised to play a crucial role in enhancing recommenders with the incorporation of LLMs.

\subsection{Self-Supervised Learning for Dynamic Recommendation Adaptation}
Current recommendation research often assumes a fixed number of users and items, making it challenging for methods to adapt to continuous new data. In practice, models are deployed in dynamic environments with continuously generated user-item interactions and new users/items~\cite{you2022roland}. To address this, prompt tuning~\cite{lester2021power, wei2024promptmm} is utilized to efficiently update pre-trained models on new data~\cite{yang2023graph}. Nevertheless, new data sparsity hinders effective supervision signals. Thus, effectively leveraging self-supervised learning for efficient learning on dynamic data is crucial.

\subsection{Theoretical Foundation of various SSL paradigm}
While self-supervised learning has improved recommender systems, a comprehensive theoretical foundation for various paradigms is lacking. Some methods have provided theoretical explanations for contrastive learning~\cite{SGL, DCCF} and generative masked autoencoding in graphs~\cite{MaskGAE}. However, recent paradigms like denoised diffusion still need theoretical explanations to demonstrate their benefits in recommendation. This understanding would help uncover the underlying principles and mechanisms driving self-supervised learning in recommendation systems and offer insights into the generalization and robustness properties of these algorithms.

%% file: conclusion.tex
\section{Conclusion}
\label{sec:conclusion}

This survey systematically reviews self-supervised learning (SSL) frameworks for recommender systems, addressing data sparsity challenges and enhancing predictive accuracy. Through analysis of hundreds of studies, we establish a comprehensive taxonomy categorizing SSL paradigms including contrastive, generative, and adversarial learning. Our findings highlight SSL's transformative role in improving personalization and user experience in recommendation systems, serving as a foundational resource for researchers developing state-of-the-art recommender systems.

\section*{Acknowledgments}
This work is supported by the National Natural Science Foundation of China under Grants 624B2122.